\documentclass[12pt,english,american]{article}
\usepackage[T1]{fontenc}
\usepackage{a4wide}
\usepackage{amsmath}
\usepackage{graphicx}
\usepackage{setspace}
\makeatletter

\providecommand{\LyX}{L\kern-.1667em\lower.25em\hbox{Y}\kern-.125emX\@}

 \newenvironment{lyxlist}[1]
   {\begin{list}{}
     {\settowidth{\labelwidth}{#1}
      \setlength{\leftmargin}{\labelwidth}
      \addtolength{\leftmargin}{\labelsep}
      }}
   {\end{list}}

\usepackage{babel}
\makeatother
\begin{document}

\title{Energy and spatial resolution of a large volume liquid scintillator
detector}

\author{O.Ju.Smirnov}

\maketitle
The result of studies of the energy and spatial resolution of a large
volume liquid scintillator detector are presented. The relations are
obtained, which allows an estimation of the detector's energy and
spatial resolutions without modeling. The relations are verified with
the data obtained with the CTF detector, which is the prototype of
the new detector for solar neutrinos, called Borexino. An event reconstruction
technique using charge and time data from the PMTs is analyzed using
the obtained relations in order to improve the detector resolution.
Algorithms for the energy and coordinates reconstruction of events
are presented\footnote{Published in:
    \\Instrum.Exp.Tech. 46 (2003) 327-344
    \\Prib.Tekh.Eksp. 2003 (2003) no.3, 49-67 (in Russian)
    \\DOI: 10.1023/A:1024458203966}.

\newpage
\section*{The list of the notations and abbreviations used in the article:}

\begin{lyxlist}{00.00.0000}
\item [$N_{PM}$]total number of the PMTs of the detector; 
\item [$Q$]total charge registered by the detector; 
\item [$\mu _{0}=\frac{Q}{N_{PM}}$]mean charge registered by the one PMT
of the detector; 
\item [$\mu _{i}$]the mean charge registered by the $i$--th PMT of the
detector; 
\item [$s_{i}=\frac{\mu _{i}}{\mu _{0}}$]relative sensitivity of the $i$--th
PMT;
\item [$v_{1}=(\frac{\sigma _{\mu _{1}}}{\mu _{1}})^{2}$]relative variance
of the single photoelectron spectrum;
\item [$f(\overrightarrow{r})$]light collection function of the detector
that relates the charge registered by a PMT for the source at position
with coordinates $\overrightarrow{r}$ to the charge, registered by
the same PMT for the same source positioned at the detector's center.
Coordinates $\overrightarrow{r}$ are the source coordinates in the
PMT coordinate system;
\item [$f_{s}(\overrightarrow{r})$]light collection function of the detector
that relates the charge registered by a detector for the source at
position with coordinates $\overrightarrow{r}$ to the charge, registered
by the detector for the same source positioned at the detector's center.
Coordinates $\overrightarrow{r}$ are the source coordinates in the
detector coordinate system;
\item [\textbf{PMT}-]photoelectron multiplier tube;
\item [\textbf{CTF}-]counting test facility of the Borexino detector;
\item [\textbf{SER}-]single electron response (charge spectrum corresponding
to a single photoelectron);
\item [\textbf{p.e}.-]photoelectron.\newpage

\end{lyxlist}

\section{Introduction}

The energy resolution of small volume scintillator detectors has been
studied during the early years of the scintillation detector development.
A typical volume of the scintillators used was no more than some litres.
A good review of the scintillation technique can be found in \cite{Birks}.
The spatial and energy resolutions of recently constructed large volume
liquid scintillator detectors is usually being studied with Monte-Carlo
simulations (see i.e. \cite{Borex},\cite{ORION}). The size of such
detectors varies from some $m^{3}$ (CTF \cite{CTF}, ORION \cite{ORION}),
up to hundreds (Borexino, \cite{Borex}), and even thousands of cubic
meters (KamLand, \cite{KamLand}).

In large volume liquid scintillator detectors scintillation is registered
by the large number of the photomultiplier tubes (PMTs) uniformly
distributed around the active volume of the detector. For each event
the charge and time of the signal arriving is measured at each PMT
of the detector. On the base of this information the energy and position
of each event is reconstructed and used in further analysis. In the
present article some relations are obtained that provide numerical
estimations of the energy and spatial resolution of a large volume
liquid scintillator detector without Monte Carlo simulations. The
relations are obtained for the case of a detector with a spherical
symmetry, but can be easily generalized for the other detector geometries.
The data of the CTF detector \cite{CTF} are used to check the validity
of the estimations. The CTF detector was operating during the years
1995-1996, its upgrade is now being used for the scintillator purity
tests for the Borexino detector.

\section{Light collection function}

In this section the light collection functions of the detector are
defined and estimated using the CTF data. The CTF consists of 4.3
tones of liquid scintillator, contained in a transparent spherical
inner vessel with a diameter of 105 cm, and viewed by 100 photomultipliers
(PMTs) located on a spherical steel support structure . The PMTs are
equipped with the light concentrator cones to increase the light collection
efficiency; the total geometrical coverage of the system is $21\%$.
The radius of the sphere passing through the opening of the light
cones is 273 cm. The detailed description of the CTF detector can
be found in \cite{CTF}. The study of the light propagation in a large
volume liquid scintillator detector was a part of the CTF programme
\cite{LightPropagation}.

The light propagation in the scintillator is usually studied in small
size laboratory facilities. Typically, it is a system consisting of
the light source of a certain wavelength (laser), a transparent cell
filled with scintillator, and a light sensor. In the laboratory studies
performed in Borexino programme the sample size was from some $cm^{3}$
\cite{ScintProp2}, to some liters \cite{LightPropagation}. In these
studies the scintillator was selected that best fits the Borexino
requirements. The probability of light absorption and reemission was
defined, as well as the probability of photon elastic scattering,
which is of a high importance due to the fact that the length of scattering
is comparable to the size of the detector. The parameters obtained
in the laboratory measurements have been used later in the Monte Carlo
simulations of the detector response.

The CTF programme included a set of measurements with a $^{222}Rn$
source with the purpose of studying the light propagation in the scintillator.
The source consisted of a $^{222}Rn$ spiked scintillator contained
in a quartz vial, which could be inserted into the detector\cite{222Radon}.
The $\beta $-decay of $^{214}Bi$ of the $^{222}Rn$ decay sequence
followed by the $\alpha $-decay of $^{214}Po$ with a mean lifetime
of $236\: \mu s$ was used to select {}``radon events'' (the events
of the $^{214}Po$ decay). The amount of light emitted in an $\alpha $-decay
of $^{214}Po$ corresponds to 862$\: keV$ energy deposited by an
electron. The obtained data were used to calibrate the event reconstruction
algorithm.

\subsection{Definitions}

\subsubsection{Light collection function for the one PMT of the detector }

\begin{figure*}

\caption{\label{Fig:CoordSystem}The coordinate system of the detector (axes
X,Y and Z) and of the PMT (axis Z$_{i}$ and angle $\Theta _{i}$).
The event occurs at the point with coordinates $\protect\overrightarrow{r}$
in the coordinate system of detector. The Z$_{i}$-axis passes from
the detector's center to the PMT, the angle $\Theta _{i}$ is calculated
from Z$_{i}$. Because of the detector's spherical symmetry the event
in the $i$-th PMT coordinate system is characterized by the pair
of polar coordinates $\{r,\Theta _{i}\}$.}

\begin{center}\includegraphics[  width=1.0\paperwidth,
  height=0.35\paperwidth,
  keepaspectratio]{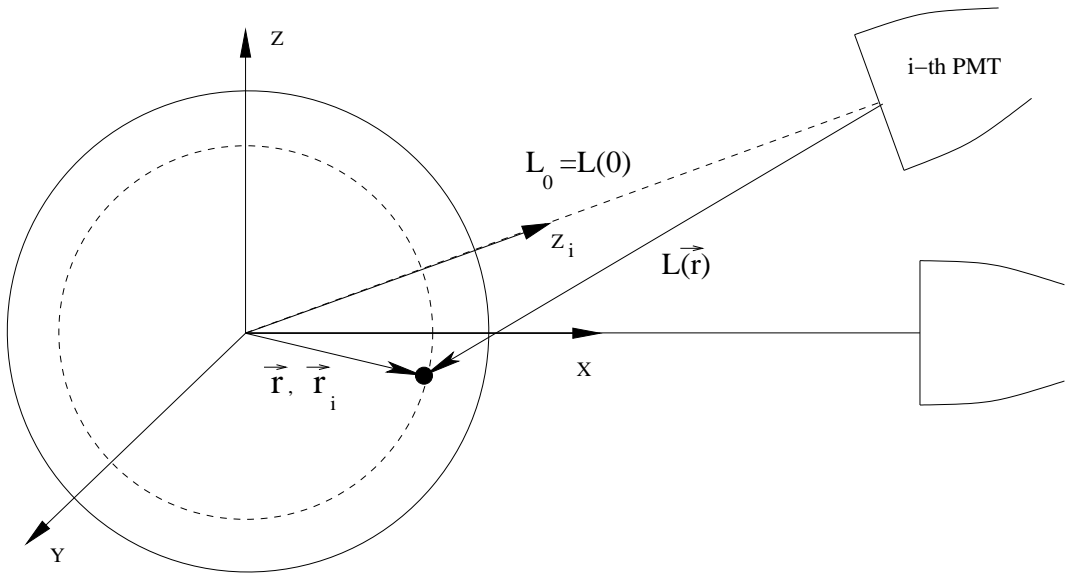}\end{center}
\end{figure*}

Let us consider a monoenergetic source at an arbitrary position in
the detector. The mean charge $\mu (\overrightarrow{r},E)$ registered
by the $i-th$ PMT for the source with energy $E$, at a point with
coordinates $\overrightarrow{r}$, can be calculated from the known
registered charge $\mu _{0}(E)\equiv \mu (\overrightarrow{0},E)$
for the same source at the detector's center using the light collection
function $f(\overrightarrow{r})$. Because of the detector's spherical
symmetry it is convenient to use a light collection function $f(\overrightarrow{r})$
in the polar coordinate system related to the PMT: the origin coincide
with a detector's geometrical center, the Z$_{i}$-axis passes from
the detector's center to the PMT as shown in fig.\ref{Fig:CoordSystem}.
Due to the detector's spherical symmetry, the light collection function
depends only on the distance from the source to the detector's center
$r$, and the polar angle $\Theta _{i}$, where the angle $\Theta _{i}$
is calculated from the Z$_{i}$-axis (see fig.\ref{Fig:CoordSystem}).
Let us define the geometrical function of one PMT as: 

\begin{equation}
f(r,\theta )\equiv \frac{\mu _{i}(\overrightarrow{r},E)}{\mu _{i}(\overrightarrow{0},E)},\label{Formula:SinglePMTfgeom}\end{equation}

where $\mu _{i}(\overrightarrow{r},E)$ is mean charge registered
by $i-$th PMT for the source with an energy $E$ placed at the point
with coordinates $\overrightarrow{r}$. Index of $\Theta _{i}$ is
omitted because of the independence of definition (\ref{Formula:SinglePMTfgeom})
on the PMT position. Let us notice that the function $f(r,\theta )$
in (\ref{Formula:SinglePMTfgeom}) doesn't depend on the PMT sensitivity. 

If the geometrical function $f(\overrightarrow{r})$, the PMT relative
sensitivities, the source coordinates and energy are known, then one
can calculate the mean registered charge for the any PMT for the source
at an arbitrary point in the detector:

\textbf{\begin{equation}
\mu _{i}(\overrightarrow{r},E)=\mu _{0}(E)\cdot s_{i}\cdot f(\overrightarrow{r_{i}})=\mu _{0}(E)\cdot s_{i}\cdot f(r,\Theta _{i})\, ,\label{fGeom}\end{equation}
}

where: 

\begin{lyxlist}{00.00.0000}
\item [$\overrightarrow{r}$,$\overrightarrow{r}_{i}$]are the coordinates
of the source in the detector's and $i-$th PMT coordinate system
respectively;
\item [$\mu _{0}(E)$]is the mean charge registered by one PMT of the detector
for the source with energy $E$ positioned at the detector's center
(averaged over all PMTs);
\item [$s_{i}\equiv \frac{\mu _{i}(E)}{\mu _{o}(E)}$]is the relative sensitivity
of the $i$-th PMT.
\end{lyxlist}
Here and below we will assume that for a point-like source at the
detector's center, the mean charge registered by the PMTs is proportional
to the source energy:

\begin{equation}
Q_{0}\equiv \sum ^{N_{PM}}\mu _{i}(0,E)=\mu _{0}N_{PM}=A\cdot E,\label{Formula:Q=AE}\end{equation}

where $A$- specific light yield measured in photoelectrons per MeV,
$E$ is the source energy and $N_{PM}$ is the number of PMTs in detector.

\subsubsection{\label{SubSection:SinglePMTlightCollectionFunction}Light collection
function for the one PMT of the CTF detector }

\begin{figure*}

\caption{\textbf{\label{Fig:CTF_Geom_Function}}The CTF single PMT light collection
function. The upper plot is the result of a MC simulation, the lower
one is obtained using CTF-I data with the radon source in different
positions. On the abscissa axis represents the distance between the
source and the detector's center, the ordinate axis represents the
polar angle $\Theta $. The range of $r$ corresponds to the radius
of the inner vessel filled with scintillator, i.e. 105 cm. The $\Theta $
angle is calculated from the axis Z passing from the detector's center
to the PMT, its range is 180 degrees.}

\begin{center}\includegraphics[  width=0.90\paperwidth,
  height=0.90\paperwidth,
  keepaspectratio]{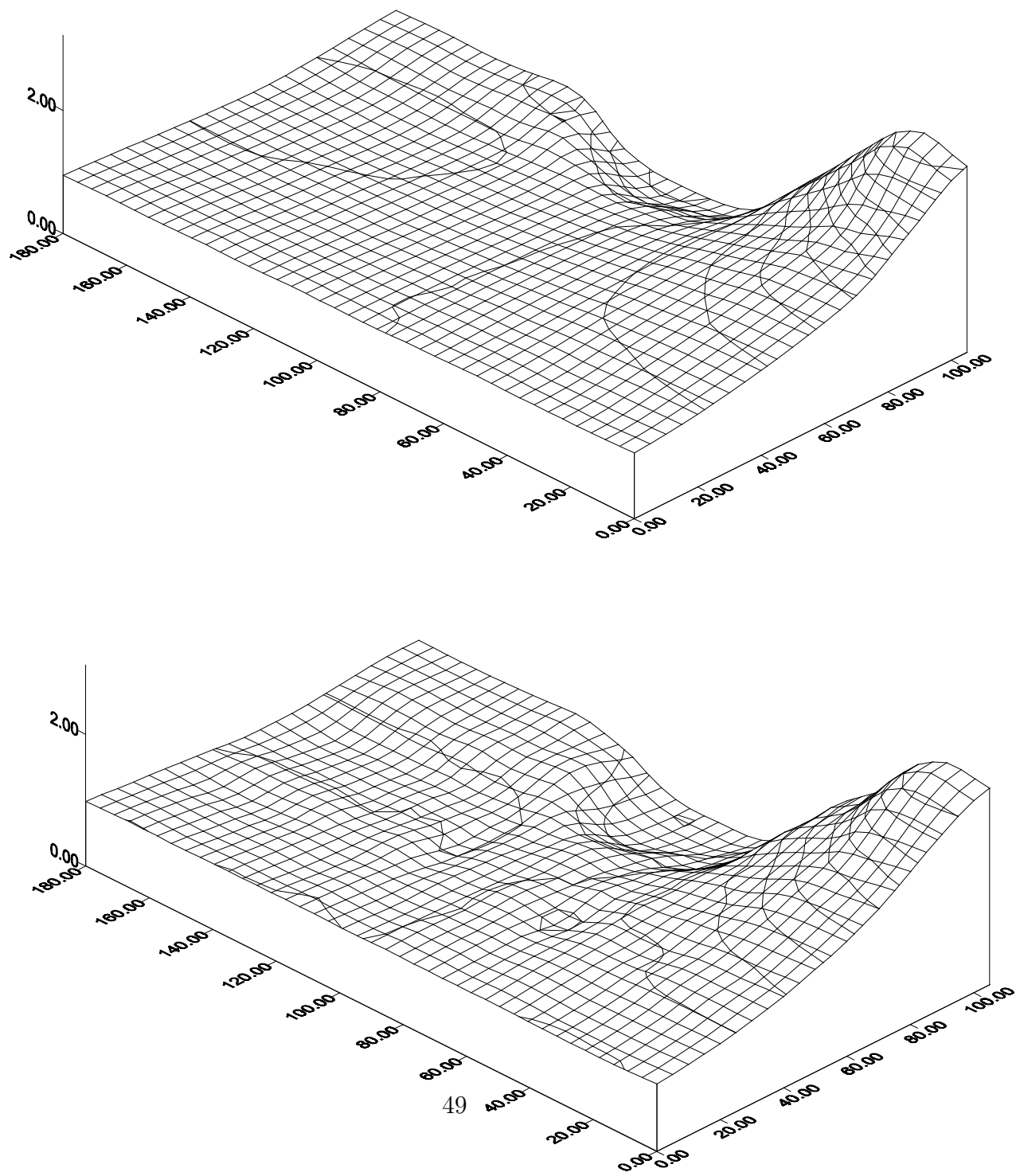}\end{center}
\end{figure*}

In the upper plot of figure \ref{Fig:CTF_Geom_Function} the light
collection function, obtained with a Monte Carlo simulation, is presented.
The light collection function is plotted in (r,$\Theta $) coordinates
of the PMT. The $r$ range corresponds to the inner vessel radius,
105 cm. The angle $\Theta $ is counted from the axis Z in the coordinates
system of the PMT\c{ } its range is from 0 to 180 degrees.

The light collection function presented in the lower plot in fig.\ref{Fig:CTF_Geom_Function}
has been obtained using data with the radon source in different positions
inside the inner vessel. Every source position gives $N_{PMT}$ points
for the \textbf{$f(r,\theta )$} function, because every PMT {}``sees''
the source in its own coordinate system. So even a restricted data
set (50 source positions have been used) allows one to follow the
light collection function over all the range of {\large r} and $\Theta $.
For the estimations, the range of {\large r} and $\Theta $ {\large }was
divided into 21x40 bins, that correspond roughly to the 5x15x15 $cm^{3}$
bins on the border of the detector's active region ($r\simeq 100$
cm). The table $\{r_{i},\Theta _{i},f(r_{i},\Theta _{i})\}$ has been
filled for every source position. After filling, the mean value at
every bin has been estimated using the number of events as statistical
weights. The empty bins were filled with the mean values of the non-empty
neighboring bins using the same weights.

\begin{figure*}

\caption{\label{Fig:ContourPlot}Isolevels for the two functions presented
in fig.2. On the abscissa axis represents the distance between the
source and the detector's center, the ordinate axis represents the
polar angle $\Theta $. }

\begin{center}\includegraphics[  width=0.90\paperwidth,
  height=0.80\paperwidth,
  keepaspectratio]{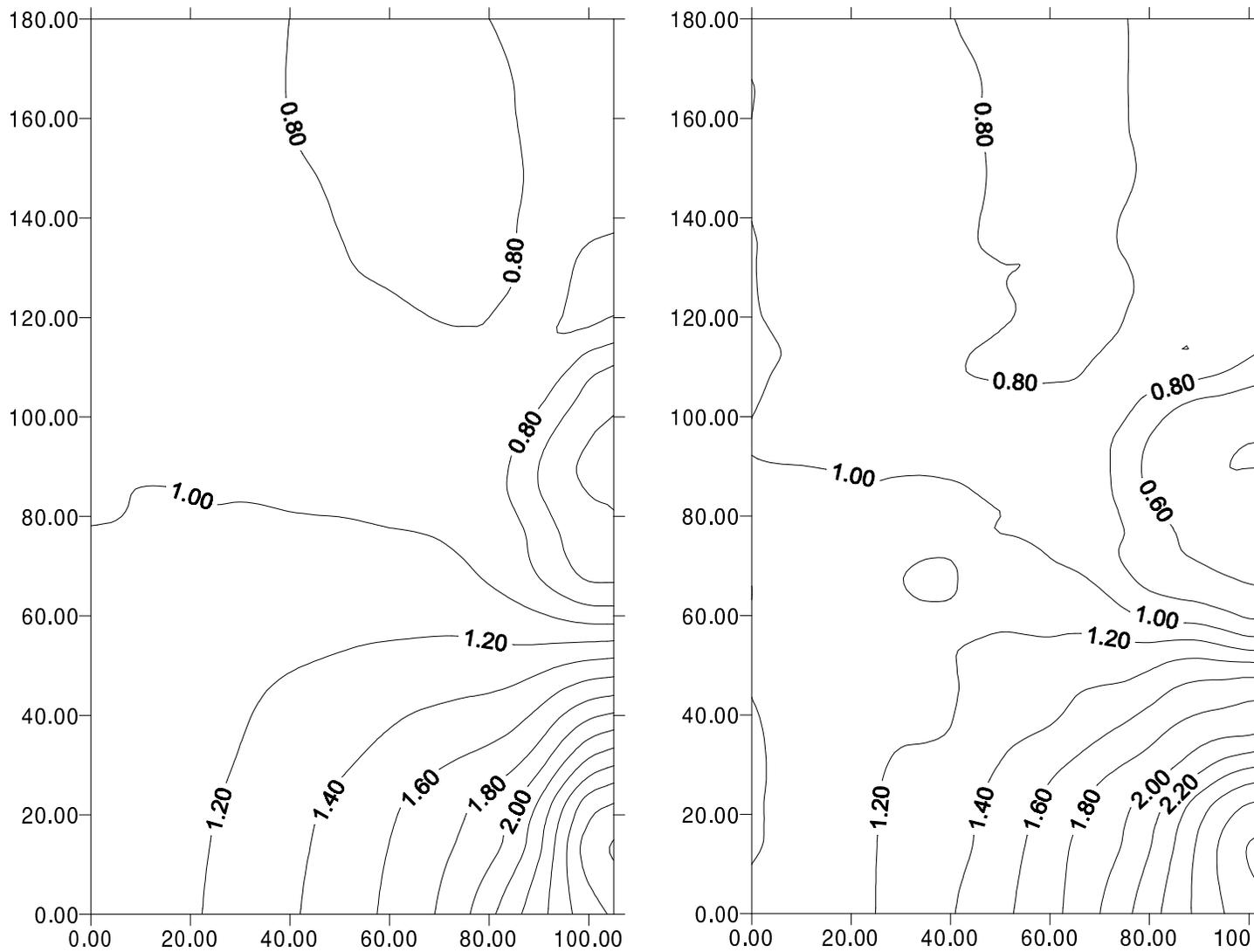}\end{center}
\end{figure*}

In fig.\ref{Fig:ContourPlot} the contour plot is presented for both
of the functions from fig.\ref{Fig:CTF_Geom_Function}. The functions
are in a good agreement, confirming the choice of the model for the
description of the detector. For large distances of the source from
the detector's center, one can see the characteristic features of
the light collection function: the {}``blind'' region in the $r>90$
cm and $\Theta \simeq 90^{o}$, and a {}``brighter'' region in comparison
to the pure solid angle function regions at $r>90$ cm and $\Theta \simeq 0^{o}$
($\Theta \simeq 180^{o}$). The presence of the {}``blind'' region
is a consequence of the the total internal reflection on the scintillator/buffer
liquid interface (water in the case of CTF) at $r>R_{0}\frac{n_{H_{2}O}}{n_{scint}}\simeq 93$
cm, where $\frac{n_{H_{2}O}}{n_{scint}}$ is the ratio of the refraction
indexes of water and scintillator. The influence of the total internal
reflection is noticeable already at $r\approx 80$$\: $cm in the
real data. The reason is the deviation of the surface shape from the
ideal sphere; where the strings holding the inner vessel are deforming
the surface.

\subsubsection{The simplified light collection function}

For numerical estimation of the detector's resolutions let us use
a simplified light collection function of the detector, preserving
only solid angle dependence and neglecting the light absorption and
reemission effects, as well as the boundary effects (reflection and
refraction on the scintillator/buffer liquid interface). The influence
of the boundary effects in any case is negligible for the events close
to the detector's center. The simple light collection function has
a following form:

\textbf{\begin{equation}
f(\overrightarrow{r})=\frac{L^{2}(0)}{L^{2}(\overrightarrow{r})}\cos \Theta \, ,\end{equation}
}

where $L(\overrightarrow{r})$ is the distance between the source
and the PMT, and $\Theta $ is the angle of incidence of light on
the PMT. From elementary geometrical considerations (see fig.\ref{Fig:CoordSystem})
one can easily obtain: 

\textbf{\begin{equation}
f(\overrightarrow{r})=\frac{L_{0}^{2}}{L^{3}(\overrightarrow{r})}\left(L_{0}-r\cdot \cos (\Theta _{0})\right)\, ,\label{Eq:SimpleGeomFunction}\end{equation}
}

where $\Theta _{0}$ is polar angle of the event in the PMT's coordinate
system and $L_{0}\equiv L(0)$.

A better approximation of the light collection function can be obtained
taking into account the light absorption in the scintillator. If,
for an event with coordinates $\overrightarrow{r}$, the path of light
in the scintillator is $L_{1}($$\overrightarrow{r}$), then the simplified
light collection function will have the following form: 

\textbf{\begin{equation}
f(\overrightarrow{r})=\frac{L_{0}^{2}}{L^{3}(\overrightarrow{r})}\left(L_{0}-r\cdot \cos \Theta _{0}\right)\cdot \exp \left(-\frac{L_{1}(\overrightarrow{r})}{L_{A}}+\frac{R_{Det}}{L_{A}}\right)\, ,\label{Eq:SimpleGeomFuncWithAbs}\end{equation}
}

where $R_{det}$ is the detector radius. The exponential factor is
equal to 1 at the detector's center ($L_{1}(0)=R_{det}$ by definition).

\subsubsection{Light collection function of the detector}

The light collection function of the detector can be defined in the
same way as the light collection function (\ref{Formula:SinglePMTfgeom})
for one PMT. It is the ratio of the total collected charge $Q(\overrightarrow{r})$
registered at a point with coordinates $\overrightarrow{r}$, to the
total collected charge registered at the detector's center for an
event of the same energy:

\begin{equation}
f_{s}(\overrightarrow{r})\equiv \frac{Q(\overrightarrow{r})}{Q_{0}}.\label{Def:f_s}\end{equation}

Replacing in (\ref{Def:f_s}) the total collected charge $Q(\overrightarrow{r})$
by the sum over all PMTs of the detector, and taking into account
the PMT relative sensitivities, one can obtain a relation linking
the light collection function $f_{s}(r)$ of the detector with the
light collection function of the PMT $f(\overrightarrow{r})$ :

\begin{equation}
f_{s}(r)\equiv \frac{1}{N_{PM}}\sum _{i}^{N_{PM}}s_{i}\cdot f(\overrightarrow{r_{i}}).\label{fs_r}\end{equation}

The light collection function $f_{s}(r)$ of the detector describes
the source position dependence of the total charge registered by a
detector. Because of the detector's spherical symmetry, the function
$f_{s}$ depends only on the distance from the source to the detector's
center. Noting that expression (\ref{Def:f_s}) is the mean of the
product of the independent quantities, one can write:

\begin{equation}
f_{s}(r)\simeq \frac{1}{N_{PM}}\sum _{i}^{N_{PM}}s_{i}\cdot \frac{1}{N_{PM}}\sum _{i}^{N_{PM}}f(\overrightarrow{r_{i}})\simeq \frac{1}{2}\int _{0}^{\pi }f(r,\Theta )\sin (\Theta )\, d\Theta \: .\label{f_s_r}\end{equation}
Here an approximate equality is used to underline the approximate
nature of the passing from the summation over PMTs to the integration
of the continuous function. The relation $<s>=\frac{1}{N_{PM}}\sum _{i}^{N_{PM}}s_{i}=1$,
coming from the definition of the relative sensitivity of the PMTs,
was used to obtain (\ref{f_s_r}). 

The assumption of the ideal spherical symmetry of the detector needs
verification, because of the different sensitivities of the PMTs and
the nonuniform PMT distribution over the spherical surface. The modeling
of the CTF detector shows that up to $r=60$ cm the equality is satisfied
within the precision of the calculations. At bigger $r$ the deviation
does not exceed $1-2\%$ percent.

\subsubsection{\label{SubSection:CTFlightCollection}CTF light collection function }

\begin{figure*}

\caption{\label{Fig:f_s}The dependence of the detector light collection function
$f_{s}$ on the source distance from the detector's center. The values
of the $f_{s}$ function calculated for the real source positions
are shown by stars. Every dot corresponds to the value calculated
at center of each 10x10x10 $cm^{3}$ bin using experimental data. }

\begin{center}\includegraphics[  width=1.0\paperwidth,
  height=0.50\paperwidth,
  keepaspectratio]{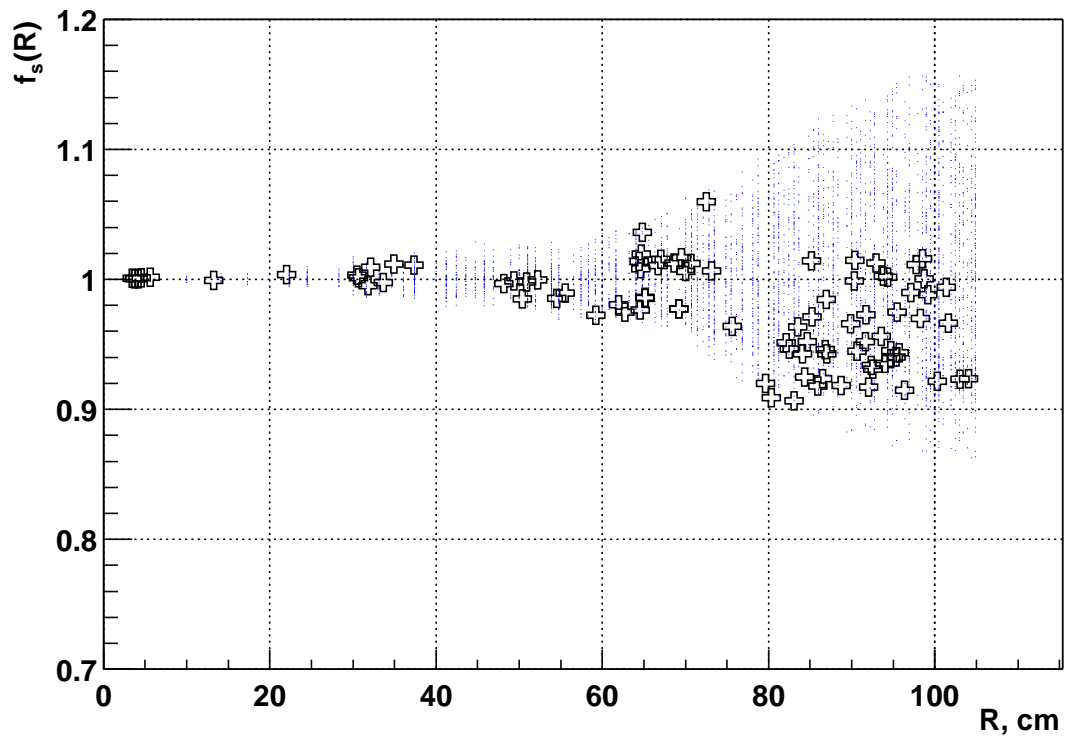}\end{center}
\end{figure*}

The dependence of the $f_{s}(\overrightarrow{r})$ function for CTF
on the source distance from the detector's center is plotted in fig.\ref{Fig:f_s}.
The volume of the detector has been divided into 10x10x10 $cm^{3}$
bins. The value of the $f_{s}$ function was calculated for each bin,
as shown by dots in scatter plot$\: $\ref{Fig:f_s}. The values of
the $f_{s}(\overrightarrow{r})$ function calculated for the nominal
source positions are shown on the same plot (black circles). The obtained
light collection function was used in the event reconstruction algorithm
which is described in section \ref{Section:DetSpatialResolutions}.

\begin{figure*}

\caption{\label{Fig:Charge}The total charge registered for the different
source positions in CTF (crosses). The values of the total charge
for the same source positions corrected with the $f_{s}$ function
are marked with circles.}

\begin{center}\includegraphics[  width=1.0\paperwidth,
  height=0.50\paperwidth,
  keepaspectratio]{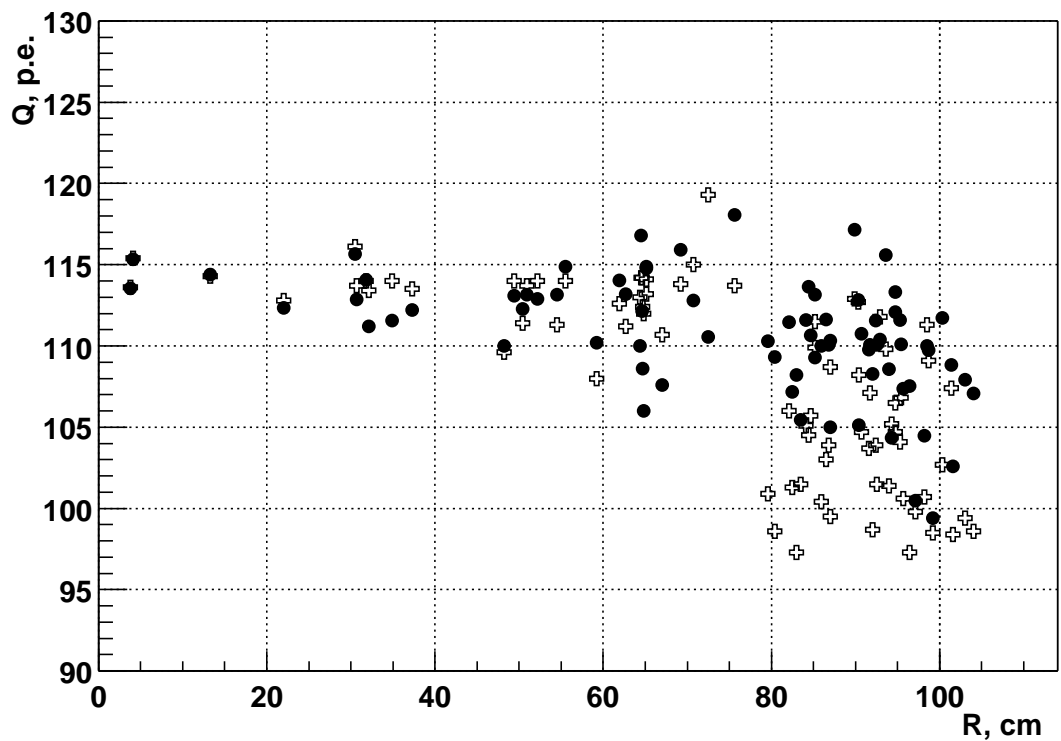}\end{center}
\end{figure*}

The strong dependence of the total collected charge on the source
position was observed in the CTF data (fig.\ref{Fig:Charge}). The
mean value of the total collected charge, and its variance over all
source positions is $107.9\pm 5.8\, (5.4\%)\: p.e.$ On the same plot
are shown the values of the total charge for the same source positions
corrected with the $f_{s}$ function (\ref{fs_r}). The mean value
and its variance over all source positions is $111.0\pm 4.4\, (3.9\%)\: p.e.$
One can see that the correction of the data with the $f_{s}$ function
improves the estimation of the energy using relation (3).

\subsection{\label{Sec:Integrals}Integrals of the CTF light collection functions}

Integrals of the light collection function can be used for a check
of the evaluation of the light collection function. The integrals
that can be calculated directly from the experimental data are of
interest: the mean value of the light collection function of the PMTs
over the detector volume $f_{V}\equiv <f(\overrightarrow{r})>_{V}$,
the mean value of the light collection function of the detector $f_{s}(\overrightarrow{r})$
over the detector volume $<f_{s}>_{V}$, and their relative variances:

\begin{equation}
v(f)\equiv \frac{<f^{2}(\overrightarrow{r})>_{V}-<f(\overrightarrow{r})>_{V}^{2}}{<f(\overrightarrow{r})>_{V}^{2}}\label{Def:v(f)}\end{equation}

\begin{equation}
v(f_{s})\equiv \frac{<f_{s}^{2}(r)>_{V}-<f_{s}(r)>_{V}^{2}}{<f_{s}(r)>_{V}^{2}}\label{Def:v(fs)}\end{equation}

Using the definitions of the light collection function of PMT (\ref{fGeom})
and light collection function of detector (\ref{Def:f_s}), one can
easily check that $f_{V}=<f(\overrightarrow{r})>_{V}=<f_{s}>_{V}$
and $v(f)\neq f(f_{s})$.

The most appropriate data for the evaluation of the integrals of the
light collection functions are the events from the radiative decay
of the isotopes from the decay chain of the U/Th. These events are
uniformly distributed over the detector volume and well identified
by the energy and time correlations. The integrals obtained from the
experimental data should coincide with the results of the numerical
integration of the light collection functions obtained from the measurements
with a point-like source or by the Monte-Carlo modeling.

The measurements with a radon source in the CTF detector were used
to calculate the light collection function of a single PMT in the
CTF detector (see subsection \ref{SubSection:SinglePMTlightCollectionFunction}).
The integrals of the light collection function were calculated using
radon decay events uniformly distributed over the detector's volume.
These data were collected during the initial stage of the CTF operation,
when a significant amount of atmospheric radon was introduced in the
detector's scintillator \cite{CTF}. The total number of 7000 radon
events was registered. The method of estimation of the integrals of
the light collection function is described later (see \ref{Section:Integrals}). 

The mean value of the $f_{s}$ function over the detector volume is
$<f_{s}(r)>_{V}=<f(r)>_{V}0.98$. The relative variance of the CTF
light collection function was $v(f_{s})=0.0023.$

The mean quadratic value of the single PMT light collection function
is: $<f^{2}(r)>_{V}=1.16$, which corresponds to the variance of the
light collection function for a single PMT $v(f(r))=0.16.$ This should
be taken into account when estimating the PMT parameters using the
events distributed over the detector's volume.

\section{Energy resolution}

If the amount of light emitted in a scintillation event is normally
distributed, then the relative variance of the PMT signal can be expressed
using the single photoelectron response $v_{1}=\left(\frac{\sigma _{\mu _{1}}}{\mu _{1}}\right)^{2}$
of the PMT, and the mean number of photoelectrons (p.e.) $\mu $ registered
in a scintillation event (see i.e. \cite{Birks}):

\begin{equation}
v\equiv \left(\frac{\sigma _{\mu }}{\mu }\right)^{2}=\frac{1+v_{1}}{\mu }.\label{var}\end{equation}

The description of the methods of PMT calibration in the single photoelectron
regime and the practical ways to define parameter $v_{1}$ one can
find in \cite{Filters}.

The physical meaning of equation (\ref{var}) is straightforward.
If $v_{1}\rightarrow 0$ (as for a delta-function response of a PMT)
then the relative variance of a PMT response is that of a Poisson
distribution ($\frac{1}{\mu }$) for the mean number of registered
p.e. $\mu $. This relation gives a fundamental limit for the PMT
energy resolution. Sometimes the deviation of the law of the p.e.
registering with the real detector from the Poisson distribution are
characterized using the so called Excess Noise Factor (ENF). Equation
(\ref{var}) will have the form $v=\frac{ENF}{\mu }$ in this case.
We will use the original equation (\ref{var}) because it corresponds
to the physics of the photoelectrons registering. Let us notice also
that, the PMT energy resolution is frequently characterized in literature
(and by manufacturers) by the peak-to-valley ratio. This parameter
is rather qualitative and can't be used for direct numerical estimations.
Indeed, two PMTs with the same peak-to-valley ratio can have very
different different single photoelectron response distributions. The
PMTs with a higher value of $v_{1}$ will demonstrate worse resolution. 

More detailed considerations of the practical scintillation counter
lead to different equations for the PMT signal relative variance,
such as (\cite{Birks}):

\begin{equation}
v=v_{p}+\frac{1+v_{1}}{\mu }.\label{varp}\end{equation}

Here the parameter $v_{p}$ is characterizing the spatial nonuniformity
of the light collection. In the proper statistical treatment this
parameter arises from averaging the probabilities of different photons
to be registered on the photocathode. The probability of a photon
registering depends on a variety of the factors, such as wavelength,
the flight path, angle and point of incidence on the photocathode,
the place in the detector where an interaction occurs, etc. The corresponding
mean number of the registered photoelectrons should be averaged over
all of these factors. This leads to the constant term in (\ref{varp}),
and sets the practical limit on the energy resolution at higher energies.
Indeed, from (\ref{varp}) it follows that $R_{max}=\frac{\sigma _{\mu }}{\mu }\rightarrow \sqrt{v_{p}}$
at $\mu \rightarrow \infty $.

In the present section the equations for the energy resolution of
a scintillator detector with a large number of PMTs will be obtained,
similar to (\ref{var}) and (\ref{varp}).

\subsection{The energy resolution for a monoenergetic point-like source}

Let us consider a source at an arbitrary position $\overrightarrow{r}$
in the detector. The mean charge registered by the $i-th$ PMT can
be calculated from the known registered charge for the same source
at the detector's center, using the light collection function (\ref{fGeom}).
The total signal of the detector is the sum of the PMT signals which
are the independent random values. Hence the mean value $Q(\overrightarrow{r})$
of the total signal is the sum of the mean values for all the PMTs,
and the variance $\sigma _{Q}^{2}(\overrightarrow{r})$ is the sum
of the variances for all the PMTs. Taking into account (\ref{fGeom}),
(\ref{Def:f_s}) and (\ref{var}) one can obtain for the total mean
charge and its variance: 

\begin{equation}
Q(\overrightarrow{r})=Q_{0}\cdot f_{s}(\overrightarrow{r})\: ,\label{Q_r_mean}\end{equation}

\begin{equation}
\sigma _{Q}^{2}(\overrightarrow{r})=Q_{0}\cdot f_{s}(\overrightarrow{r})+Q_{0}\frac{1}{N_{PM}}\sum _{i}^{N_{PM}}s_{i}\cdot f(\overrightarrow{r_{i}})\cdot v_{1_{i}}\: ,\label{SigmaSq_r}\end{equation}

where $v_{1_{i}}$ is the relative variance of the single photoelectron
response of the $i-th$ PMT.

Let us introduce a parameter of the detector $v_{1}^{Det}(\overrightarrow{r})$:

\begin{equation}
v_{1}^{Det}(\overrightarrow{r})=\frac{1}{N_{PM}}\sum _{i}^{N_{PM}}s_{i}\cdot f(\overrightarrow{r_{i}})\cdot v_{1_{i}}\: .\label{v1_Det_r}\end{equation}

Using (\ref{v1_Det_r}) from (\ref{Q_r_mean}) and (\ref{SigmaSq_r})
one can obtain the relative resolution at an arbitrary detector point:

\begin{equation}
R(Q,\overrightarrow{r})\equiv \sqrt{\frac{\sigma _{Q}^{2}(\overrightarrow{r})}{Q^{2}(\overrightarrow{r})}}=\sqrt{\frac{1+\frac{1}{f_{s}(\overrightarrow{r})}v_{1}^{Det}(\overrightarrow{r})}{f_{s}(\overrightarrow{r})\cdot Q_{0}}}.\label{ResDet_r}\end{equation}

If the number of PMTs is large enough, then the mean value of the
product in the definition of the parameter $v_{1}^{Det}(\overrightarrow{r})$
can be substituted by the product of the mean values (a check with
the CTF data shows that at $N_{PM}\ge 50$ the deviation from the
precise value is less than 1\%):

\begin{equation}
v_{1}^{Det}(\overrightarrow{r})\approx \frac{1}{N_{PM}}\sum _{i}^{N_{PM}}f_{s}(\overrightarrow{r})\cdot \frac{1}{N_{PM}}\sum _{i}^{N_{PM}}s_{i}\cdot v_{1_{i}}=f_{s}(\overrightarrow{r})\cdot v_{1}^{Det}(0)\: .\label{v1_simple}\end{equation}

Then the relative energy resolution of the detector for a point-like
source at an arbitrary point is: 

\begin{equation}
R(Q,\overrightarrow{r})=\sqrt{\frac{1+v_{1}^{Det}(0)}{f_{s}(\overrightarrow{r})\cdot Q_{0}}}.\label{ResCTF_r}\end{equation}

As it has been pointed out before, because of the detector's symmetry,
the geometrical function $f_{s}$ depends mainly on the distance from
the source to the detector's center $r$. So the energy resolution
in turn depends mainly on $r$ as well :

\begin{equation}
R(Q,\overrightarrow{r})\approx R(Q,r)=\sqrt{\frac{1+v_{1}^{Det}(0)}{f_{s}(r)\cdot Q_{0}}}.\label{Res_Det_rad}\end{equation}

The relative energy resolution $R(Q,0)$ for the point-like monoenergetic
source at the detector's center ($f_{s}(0)=1$) 

\begin{equation}
R(Q,0)\equiv \sqrt{\frac{\sigma _{Q_{0}}^{2}}{Q_{0}^{2}}}=\sqrt{\frac{1+v_{1}^{Det}(0)}{Q_{0}}}\label{ResCTF_0}\end{equation}

has the same form as for the relative energy resolution of a single
PMT (\ref{var}), with the parameter $v_{1}$ replaced by the average
parameter $v_{1}^{Det}(0)$ that also takes into account the different
relative sensitivities of the PMTs.

\subsection{The energy resolution for a non point-like source}

It will be shown that the simple law (\ref{Res_Det_rad}) does not
describe the energy resolution for events uniformly distributed over
the detector volume. The reason is that the distribution of the number
of registered photoelectrons does not follow the Poisson law in this
case.

\subsubsection{\label{Section:NonPointLikeSource}A monoenergetic source }

Let us assume that for a point-like source with an energy $E$ at
a position $\overrightarrow{r}$, the $i$-th PMT registers a mean
number of photoelectrons \textbf{$\mu _{i}(\overrightarrow{r_{i}},E)$,}
and the amount of the registered photoelectrons follows the Poisson
law. \textbf{}Here and below we will use the coordinates system of
the PMT (see fig.\ref{Fig:CoordSystem}). 

The mean number of registered photoelectrons \textbf{$\mu _{i}(\overrightarrow{r_{i}},E)$}
can be calculated using (\ref{fGeom}). Using the definition of $f_{s}\, (r)$
(\ref{Def:f_s}) one can write:

\textbf{\begin{equation}
Q(r,E)=Q_{0}(E)\cdot f_{s}\, (r)\, .\label{Formula:4}\end{equation}
}

If a source with an energy$\: $$E$ is uniformly distributed over
the detector's volume with density $n(r)$, then the mean registered
charge is:

\textbf{\begin{equation}
<Q>_{V}=Q_{0}(E)\cdot \int _{V}f_{s}\, (r)\, n(r)\, dV\equiv Q_{0}(E)<f_{s}>_{V}.\label{formula:Q_V}\end{equation}
} 

The mean value of the detector's function $<f_{s}>_{V}$ is equal
to the mean value of the single PMT function $<f>_{V}$$\: $.

For a point-like source the distribution of registered charge is the
sum of the charge distributions for the individual PMTs, hence the
variation of the registered charge is the sum of the individual variations:

\textbf{\begin{equation}
\sigma ^{2}(r,E)=\sum _{i=1}^{N_{PM}}\sigma _{i}^{2}\, (r_{i},E)=\sum _{i=1}^{N_{PM}}\mu _{i}\, (r_{i},E)\, (1+v_{1_{i}})=Q_{0}(E)\, f_{s}(r)\, (1+\overline{v_{1}})\: .\label{Formula:5}\end{equation}
}

For a source uniformly distributed over the detector volume, the mean
of the total charge squared can be obtained by averaging the mean
values of the charge squared at every point in the detector:

\textbf{\[
<\overline{Q^{2}(r,E)}>_{V}=<\overline{Q(r,E)}^{2}+\sigma (r,E)^{2}>_{V}=\]
}

\textbf{\begin{equation}
=Q_{0}^{2}(E)<f_{s}^{2}(r)>_{V}+Q_{0}(E)(1+\overline{v_{1}})<f_{s}>_{V}.\label{formula:Q2_V}\end{equation}
}

The variation of the total charge can then be obtained using (\ref{formula:Q_V})
and (\ref{formula:Q2_V}):

\begin{equation}
\sigma _{Q}^{2}=Q_{0}^{2}(E)(<f_{s}^{2}(r)>_{V}-<f_{s}>^{2})+Q_{0}(E)(1+\overline{v_{1}})<f_{s}>.\label{Formula:SigQ2}\end{equation}

Finally, the relative charge resolution of the detector is

\textbf{\begin{equation}
R_{V}=\sqrt{\frac{\sigma _{Q}^{2}}{Q^{2}}}=\sqrt{\frac{1+\overline{v_{1}}}{Q_{0}(E)<f_{s}>_{V}}+v(f_{s})}\: .\label{R_V}\end{equation}
}Here $v(f_{s})$ is the relative variance of the $f_{s}(r)$ function
over the detector volume defined by equation (\ref{Def:v(fs)}). It
should be pointed out that $v(f_{s})$ has the same sense as $v(p)$
in formula (\ref{varp}).

\subsubsection{A source with energy spectrum $f_{E}(E)$ }

If a source is distributed over the detector's volume with density
$n(r)$, and its energy spectrum is described with a function $f_{E}(E)$
then the mean charge registered by the detector is:

\begin{equation}
<Q>=Q_{0}\cdot \int _{E>E_{th}}E\, f_{E\, }(E)\, dE\cdot \int _{V}f_{s}\, (r)\, n(r)\, dV\equiv Q_{0}<E><f_{s}>_{V}.\label{formula:Q_EV}\end{equation}

The proportionality of the registered charge $Q$ to the source energy
$E$ is assumed here. Thus, $Q(E)=A\cdot E$, where $A$ is the charge
registered for the unit energy deposition, or a photoelectron yield. 

For the source distributed over the detector's volume with an energy
spectrum $f_{E}(E)$ the mean value of the registered charge squared
can be obtained by averaging the mean quadratic values of the charge
registered over the detector's volume:

\textbf{\[
<\overline{Q^{2}(r,E)}>_{V,E}=<\overline{Q(r,E)}^{2}+\sigma (r,E)^{2}>_{V,E}=\]
}

\textbf{\begin{equation}
=Q_{0}^{2}<E^{2}><f_{s}^{2}(r)>_{V}+Q_{0}(1+\overline{v_{1}})<E><f_{s}>_{V}.\label{formula:Q2_EV}\end{equation}
}

So that the variation of the total registered charge is (using (\ref{formula:Q_EV})
and (\ref{formula:Q2_EV})):

\begin{equation}
\sigma _{Q}^{2}=Q_{0}^{2}(<E^{2}><f_{s}^{2}(r)>_{V}-<E>^{2}<f_{s}>^{2})+Q_{0}(1+\overline{v_{1}})<E><f_{s}>.\label{Formula:6}\end{equation}

and the relative variance of the detector response is:

\textbf{\begin{equation}
Var_{V,E}(Q)\equiv \frac{\sigma _{Q}^{2}}{<Q>^{2}}=\frac{1+\overline{v_{1}}}{Q_{0}(E)<f_{s}>_{V}}+v(f_{s})+v(E)+v(f_{s})v(E),\label{Eq:Res_VQ}\end{equation}
}

where $v(E)$ is the relative variance of the source energy spectrum 

\textbf{\begin{equation}
v(E)\equiv \frac{<E^{2}>-<E>^{2}}{<E>^{2}}.\label{Eq:v(E)}\end{equation}
}

\subsection{\label{Sec:CalibrationInfluence}Influence of the PMT calibration
precision on the detector energy resolution }

Let us consider the influence of the precision of the PMT calibration
on the energy resolution of the detector. The calibration of the PMT
means establishing the value $q_{1}$ corresponding to the PMT single
photoelectron response. PMT calibration methods are discussed in detail
in \cite{Filters}.

Let us call $\mu _{i}^{\prime }$ the $i$--th PMT charge, defined
applying the calibration, and $\mu _{i}$ the real charge. The mean
total charge collected by the detector is obtained summing the charge
on the individual PMTs:

\textbf{\begin{equation}
Q\prime (r)=\sum _{i}^{N_{PM}}\mu \prime _{i}(\overrightarrow{r_{i}})=Q\cdot \frac{1}{N_{PM}}\sum _{i}^{N_{PM}}c_{i}\cdot s_{i}\cdot \mu (\overrightarrow{r_{i}})\label{formulaqu_p}\end{equation}
} where the parameter \textbf{$c_{i}=\frac{\mu _{i}^{\prime }}{\mu _{i}}$}
describes the precision of the calibration.

The relative variances of $\mu $ and $\mu '$ does not depend on
a calibration used; and hence the variance of the total signal of
the detector is:

\begin{equation}
\sigma _{Q^{\prime }}^{2}(r)=Q_{0}\cdot (\frac{1}{N_{PM}}\sum _{i}^{N_{PM}}c_{i}^{2}\cdot s_{i}\cdot f(\overrightarrow{r_{i}})+\frac{1}{N_{PM}}\sum _{i}^{N_{PM}}c_{i}^{2}\cdot s_{i}\cdot f(\overrightarrow{r_{i}})\cdot v_{1_{i}}).\label{sigm}\end{equation}

So the relative energy resolution of the detector for the considered
calibration in this case is:

\textbf{\begin{equation}
R^{\prime }(Q,r)=\sqrt{\frac{\frac{1}{N_{PM}}\sum _{i}^{N_{PM}}c_{i}^{2}\cdot s_{i}\cdot f(\overrightarrow{r_{i}})+\frac{1}{N_{PM}}\sum _{i}^{N_{PM}}c_{i}^{2}\cdot s_{i}\cdot f(\overrightarrow{r_{i}})\cdot v_{1_{i}}}{Q\cdot (\frac{1}{N_{PM}}\sum _{i}^{N_{PM}}c_{i}\cdot s_{i}\cdot f(\overrightarrow{r_{i}}))^{2}}}=\frac{const}{\sqrt{Q}}\label{ResCTF_prime}\end{equation}
}

In the case when the number of PMTs is large enough (in practice large
enough are values $N_{PM}>50)$ it is possible to replace the means
of the product in (\ref{ResCTF_prime}) with the product of the means.
Taking into account that $<s>=1$ by its definition, and $<c^{2}>=<c>^{2}+\sigma _{c}^{2}$$\, $,
the formula (\ref{ResCTF_prime}) is significantly simplified:

\textbf{\begin{equation}
R^{\prime }(Q,r)=\sqrt{\frac{1+\overline{v_{1}}}{f_{s}(r)\cdot Q_{0}}}\sqrt{1+v(c)}=R(Q,r)\sqrt{1+v(c)}\, ,\label{ResCTF_p}\end{equation}
}

where $v(c)\equiv \left(\frac{\sigma _{c}}{\overline{c}}\right)^{2}$
is the relative variance of the calibration accuracy, and $f_{s}(r)$
is defined by (\ref{f_s_r}).

One can see that the detector energy resolution is quite insensitive
to the individual PMT calibration. Indeed, in accordance with (\ref{ResCTF_p}),
a moderate PMT calibration precision of $20\%$ (i.e. $\sigma _{c}=0.2$)
will cause only $2\%$ degrading of the detector's resolution.

\subsection{\label{Section:Integrals}Evaluation of the integrals of the light
collection function using events uniformly distributed over the detector
volume }

During the initial stage of the CTF operation, a significant number
of radon decays were observed in the detector scintillator \cite{CTF}.
This atmospheric radon was introduced in the scintillator during the
filling. This data has been used for the detector calibrations. Below
it is demonstrated how the same data can be used to define the light
collection function and its integrals.

The mean collected charge and its variance for the radon events selected
in the different $r$ intervals ($r$ is a distance between the point
of decay and the detector's center) are presented in table$\: $\ref{Table:Radon}.
The coordinates of the events were obtained using an event reconstruction
program. 

This data permits an estimate of the light collection function $f_{s}(r)$,
giving an additional possibility to check the Monte Carlo model and
the light collection function obtained with the artificial radon source.

As one can see from fig.\ref{Fig:f_s}, the $f_{s}\simeq 1$ up to
$r<40$ cm, hence $Q(r<40cm)\simeq Q_{0}$. The mean value of the
light collection function over the detector volume can be obtained
using the value of $Q(r<40cm)$ as $Q_{0}$ and $<Q>$ for all the
radon events ($0<r<\infty $):\begin{equation}
<f(\overrightarrow{r})>_{V}=<f_{s}(r)>_{V}=\frac{<Q>_{V}}{Q_{0}}\simeq 0.98\label{Num:VolFact}\end{equation}

The radon data in the detector region $r<40\: $cm can be used to
obtain the relative resolution at the detector's center $R_{0}=0.0742$.
The radon data distributed over all the detector volume can be used
to obtain the relative resolution for the distributed source $R_{vol}=0.0874$.
The parameter $v(p)$ can now be estimated from the simple relation
following from (\ref{R_V}): $R_{vol}^{2}=\frac{R_{0}^{2}}{<f_{s}>^{2}}+v(p)$,
which yields the value of $v(p)\simeq 0.002$.

\section{\label{Section:DetSpatialResolutions}Detector's spatial resolution }

Knowledge of the event position in low-background scintillator detectors
is necessary for the rejection of correlated background events, and
for the off-line active shielding from the external backgrounds. The
event coordinates can be reconstructed using the charge signals from
the PMTs, or using the time of arrival of the signals at the PMTs.
If the precision of reconstruction by both methods is comparable,
then full information (charge and timing) can be used for improving
the reconstruction. Below, the precision of position reconstruction
is studied for all methods mentioned.

\subsection{\label{Sec:Charge_reconst}Method of reconstruction using charge
signals}

The spatial reconstruction using charge signals can be performed using
the maximum likelihood method. The 3 coordinates and full charge $Q_{0}$,
corresponding to an event of the same energy at the detector's center,
are free parameters. The likelihood function has the following form: 

\textbf{\begin{equation}
L(x,y,z,Q_{0})=\ln \left(\prod _{i=1}^{N_{PM}}p\, (\mu (\overrightarrow{r_{i}}(x,y,z),Q_{0}),q_{i})\right)\, ,\label{Eq:MaxLikeCharge}\end{equation}
}

where $p\, (\mu (\overrightarrow{r_{i}}(x,y,z),Q_{0}),q_{i})$ is
the probability to register charge $q_{i}$ at the $i$--th PMT for
the event at a position with coordinates $\overrightarrow{r}=\{x,y,z\}$
and the total charge $Q_{0}$. Here $\overrightarrow{r}_{i}(x,y,z)$
are the coordinates of the event in the $i$--th PMT coordinate system.
Using the PMT light collection function $f(\overrightarrow{r})$ and
the relative sensitivities $s_{i}$, one can write:

\textbf{\begin{equation}
\mu (\overrightarrow{r_{i}},Q)=f(\overrightarrow{r_{i}})\cdot s_{i}\cdot \frac{Q}{N_{PM}}.\label{Eq:mu(r,qu)}\end{equation}
}

The probability to register charge q at the $i$--th PMT, if the mean
expected charge is $\mu $ , assuming a Poisson distribution for the
registered number of p.e., can be written as:

\begin{equation}
p\, (\mu ,q)=\sum _{N=0}^{N_{Max}}P(N,\mu )f_{N}(q)\, ;\label{Eq:p(mu,qu)}\end{equation}
where $P(N,\mu )$ is the probability to register N p.e. if their
mean value is $\mu $, and $f_{N}(q)$ is the probability density
function of the registering charge $q$ for $N$ p.e. The model function
$f_{N}(q$) from \cite{Filters} was used in the calculation, with
the parameters averaged over all PMTs. The parameters of the model
function for each PMT were defined during the PMT testing before installation
in the detector.

\begin{figure*}

\caption{\label{Fig:Example_of_reconst_Rad}The difference between the nominal
source position and the reconstructed one using the charge signals,
at different positions inside the detector's volume: a)$r(0,0,-40)=40$
cm, b)$r(0,0,-80)=80$ cm and c)$r(32,0,105)\simeq 105$ cm (dotted
line). For comparison in the same plot the results of reconstruction
using time signals only are shown with a solid line. One can see that
reconstruction on the charge signals works better than the reconstruction
using time signals at large r values.}

\begin{center}\includegraphics[  width=0.70\paperwidth,
  height=0.25\paperwidth]{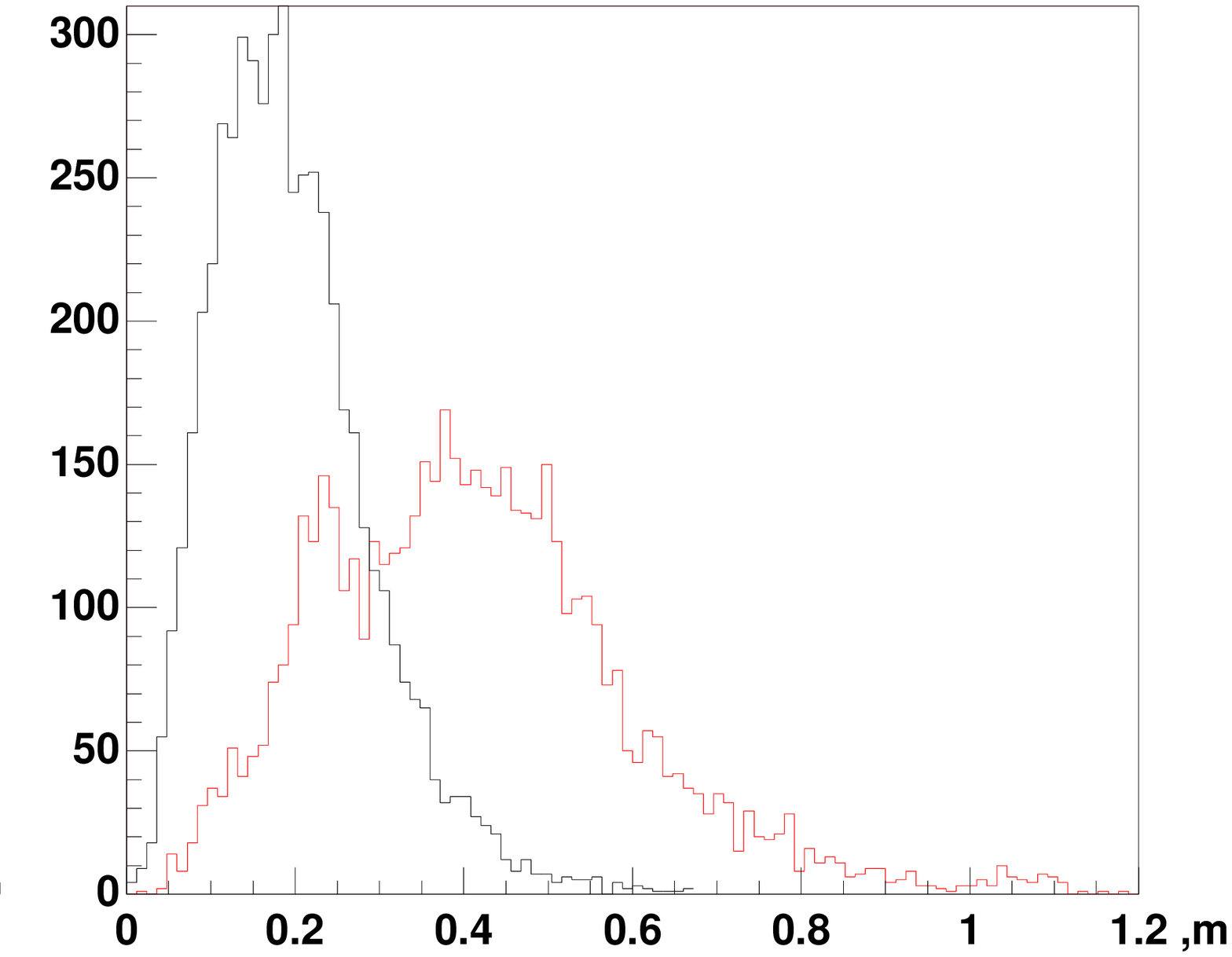}\end{center}

\textbf{a)}

\begin{center}\includegraphics[  width=0.70\paperwidth,
  height=0.25\paperwidth]{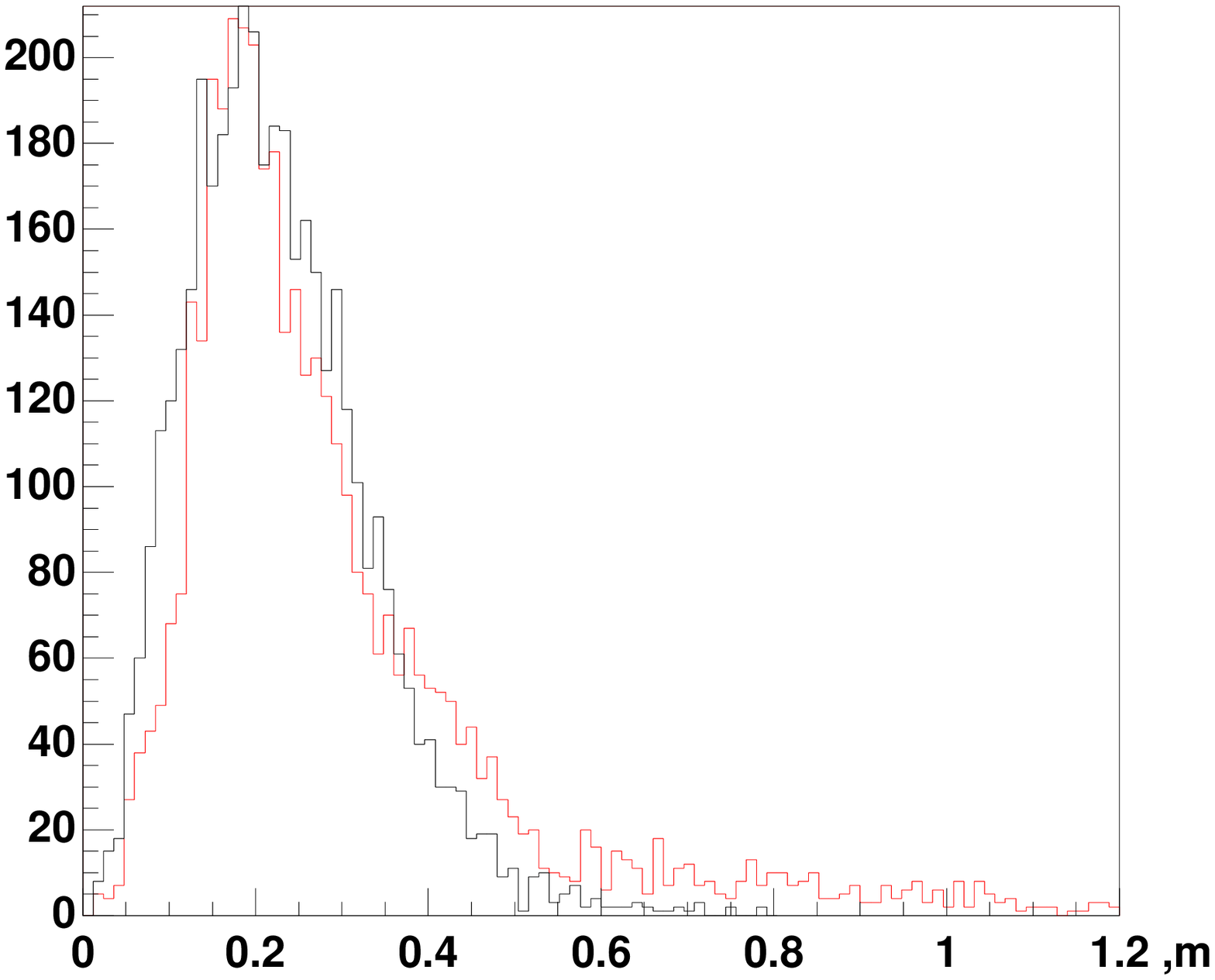}\end{center}

\textbf{b)}

\begin{center}\includegraphics[  width=0.70\paperwidth,
  height=0.25\paperwidth]{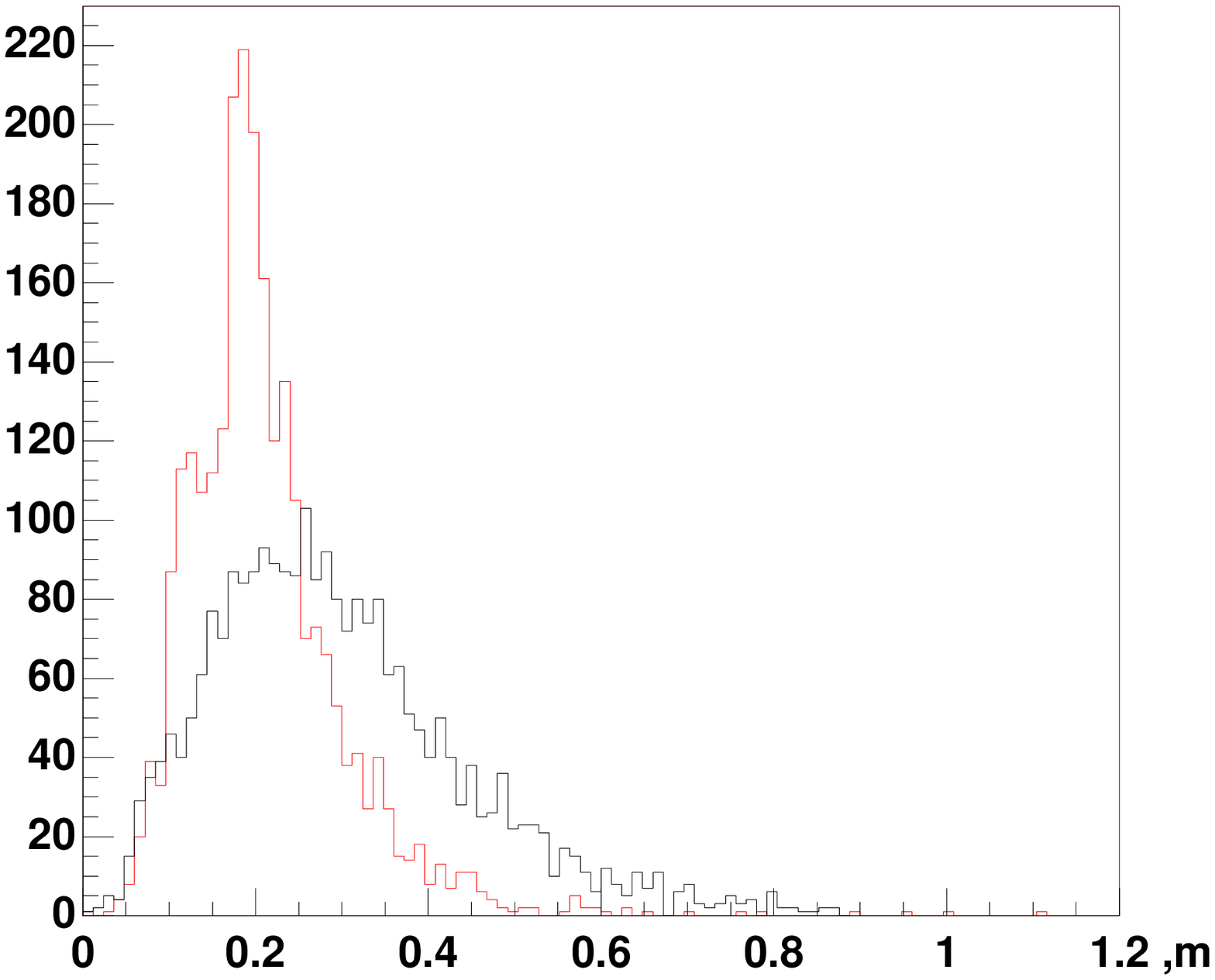}\end{center}

\textbf{c)}
\end{figure*}

The algorithm for the likelihood function construction can be divided
into the following steps:

\begin{enumerate}
\item The initial total charge value $Q_{0}$ is defined as the sum of the
charge registered at the individual PMTs. 
\item The initial coordinates $(x,y,z)$ of the event are guessed on the
basis of the signal distribution symmetries. As a first approximation
of the coordinates, the linear combination of the PMTs coordinates
with weights corresponding to the PMTs signals:\[
x_{q}=4.5\sum \frac{q_{i}x_{i}}{L_{0}\mu _{0}},\]
is used. The coefficient 4.5 was defined from the measurements with
the source at position \{0,0,94\}.
\item The mean charge $\mu (\overrightarrow{r_{i}},Q)$ expected at the
$i$--th PMT is defined using formula (\ref{Eq:mu(r,qu)});
\item Then the probability $p_{i}$ to register charge $q_{i}$ at the $i$--th
PMT for an event at a position with coordinates $\overrightarrow{r}=\{x,y,z\}$
is calculated using formula (\ref{Eq:p(mu,qu)});
\item Then the value of the likelihood function is increased by  $log(p_{i})$,
and the algorithm is repeated starting from point 3.
\end{enumerate}
Examples of reconstruction for the different radon source positions
in the CTF detector using the charge signals are shown in figure \ref{Fig:Example_of_reconst_Rad}
(dashed lines). The plot presents the distribution of the distance
from the nominal source position to the reconstructed one. For comparison
the reconstruction using the time signals is shown in the same figure
(solid lines). One can see that the reconstruction using the time
signals is better for small $r$ ($r<60$ cm). The reconstruction
with the charge signals at $r\simeq 80$ cm is comparable to the reconstruction
with the time signals, while the reconstruction with the charge signals
for the source close to the inner vessel is better than the reconstruction
with the time signals.

\subsection{Analysis of the precision of the spatial reconstruction using the
charge data}

Let us estimate the maximum possible spatial resolution when using
the charge signals for the spatial reconstruction. Let us consider
an event of an energy E at a position $r=\{x,0,0\}$. Because of the
detector's spherical symmetry this case is quite common. When the
source is moved by $\Delta x$ along the X axis the mean registered
charge will change by (see equation (\ref{fGeom})) \begin{equation}
\Delta \mu =\mu _{0}\, s_{i}\, \frac{df(r,\Theta _{i})}{dx}(x,0,0)\Delta x\: .\label{Formula:DeltaMu}\end{equation}

The uncertainty $\sigma _{\mu }$ of the charge registered by one
PMT can be obtained from (\ref{var}). If the source energy is unknown
a priori, the squared error of the registered charge reconstruction
should be added

\begin{equation}
\sigma _{q}^{2}=\frac{\sigma _{Q}^{2}}{N_{PM}}=\frac{1}{N_{PM}}\frac{1+v_{1}^{Det}}{f_{s}(r)Q_{0}}\cdot \left(f_{s}(r)\, Q_{0}\right)^{2}=\mu _{0}(1+v_{1}^{Det})\, f_{s}(r),\label{Formula:Sigq2d}\end{equation}

and the uncertainty of the coordinate reconstruction for a single
PMT is \begin{equation}
\sigma _{x_{i}}=\sqrt{f(\overrightarrow{r_{i}})\frac{1+v_{1_{i}}}{\mu _{0}}+\frac{f_{s}(r)}{s_{i}}\frac{1+v_{1}^{Det}}{\mu _{0}}}\cdot \left(\frac{df(r)}{dx}(r_{i})\right)^{-1}.\end{equation}

The signals on the PMTs are independent random values, so for the
whole detector the error of the coordinate reconstruction can be calculated
using $\sigma _{x_{i}}$: 

\begin{equation}
\sigma _{x}\geq \frac{1}{\sqrt{\sum \frac{1}{\sigma _{x_{i}}^{2}}}}=\frac{R_{0}(E)}{\sqrt{\frac{1}{N_{PM}}\left(\frac{s_{i}}{s_{i}\, f(\overrightarrow{r_{i}})\frac{1+v_{1_{i}}}{1+v_{1}^{Det}}+f_{s}(r)}\right)\left(\frac{df(r)}{dx}(r_{i})\right)^{2}}}.\label{Eq:SigmaX________}\end{equation}

It is convenient to change to the PMT coordinate system and to replace
the summing with integration:\begin{equation}
\sigma _{r}(r)\geq R_{0}(E)\left(\frac{1}{2}\int _{0}^{\pi }\frac{1}{f(r,\Theta )+f_{s}(r)}(\frac{df(r,\Theta )}{dr})^{2}\sin \Theta d\Theta \right)^{-\frac{1}{2}}.\label{DeltaX}\end{equation}

Here the differentiation over x in the detector's coordinate system
is replaced by differentiation over r in the PMT coordinate system.
In fact, $\sigma _{r}(r)$ is the detector's radial resolution. The
tangential resolution can be obtained in the same way (the factor
1/2 comes from the averaging over $\phi $ angle):

\textbf{\begin{equation}
\sigma _{\Theta }(r)\geq R_{0}(E)\left(\frac{1}{2}\cdot \frac{1}{2}\int _{0}^{\pi }\frac{1}{f(r,\Theta )+f_{s}(r)}\left(\frac{1}{r\sin \Theta }\frac{df(r,\Theta )}{d\Theta }\right)^{2}\sin \Theta d\Theta \right)^{-\frac{1}{2}}.\label{DeltaY}\end{equation}
}

Because of the detector's spherical symmetry, the radial and the tangential
resolutions define the total detector's resolution at any point in
the detector.

For estimating the detector's spatial resolution, let us use a simplified
light collection function of the detector (\ref{Eq:SimpleGeomFunction}).
This function can be easily integrated analytically. The precision
of the spatial reconstruction at the detector's center, calculated
with the function$\, $(\ref{Eq:SimpleGeomFunction}), is: \begin{equation}
\sigma ^{(q)}=\sqrt{\frac{3}{2}}L_{0}R_{0}(E)\, ,\label{SigmaXq_e}\end{equation}

where $R_{0}(E)$ is the energy resolution at the detector's center. 

A better approximation for the energy resolution at the detector's
center can be obtained by taking into account the light absorption
in the scintillator. The precision of the spatial reconstruction at
the detector's center, calculated with function (\ref{Eq:SimpleGeomFuncWithAbs}),
is: 

\textbf{\begin{equation}
\sigma ^{(q)}=\sqrt{\frac{3}{2}}L_{0}R_{0}(E)\cdot \frac{1}{1+\frac{L_{0}}{2L_{A}}}.\label{formula:SigXq}\end{equation}
}

\begin{figure*}

\caption{\label{Fig:Rex_Q_T}The spatial reconstruction precision (1$\sigma $)
using the time data (crosses) and the charge data (empty crosses)
as a function of the source distance from the detector's center (50
PMTs, CTF detector). The results of the calculation using the light
collection function estimated from the CTF data are plotted with circles.
The two lines corresponds to the calculation using the simplified
light collection function (upper solid line) and the simple light
collection function with light absorption (lower solid line). }

\begin{center}\includegraphics[  width=0.90\paperwidth,
  height=0.40\paperwidth,
  keepaspectratio]{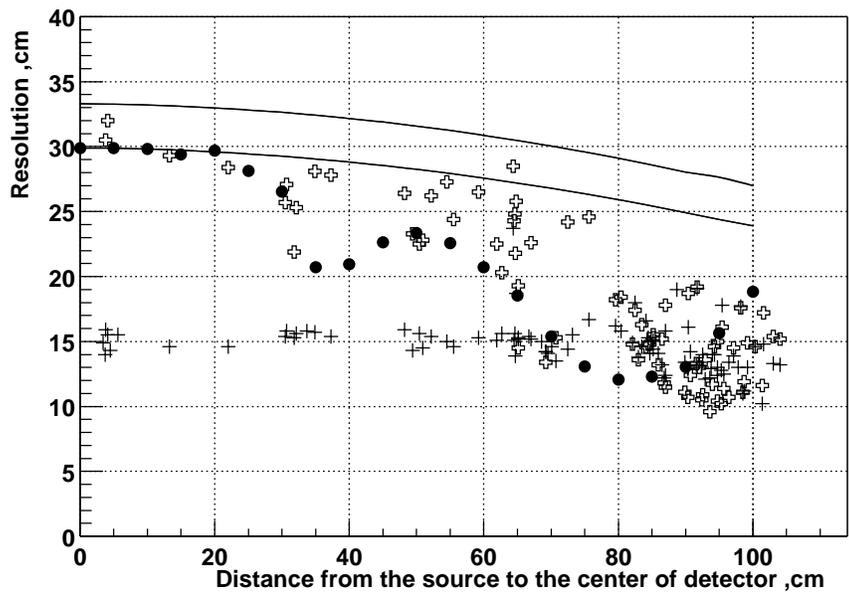}\end{center}
\end{figure*}

The comparison of the reconstruction precision for the CTF detector
using the time signals (crosses) and the charge signals (empty crosses)
for different distances from the source to the detector's center is
presented in fig.\ref{Fig:Rex_Q_T}. In order to compare the estimation
with real data, only 50 PMTs were considered (as in CTF for the runs
with the artificial radon source). In the same figure the results
of calculation of the spatial reconstruction precision are presented
for 3 different light collection functions, namely, the light collection
function obtained from the experimental data (stars), the simplified
light collection function (\ref{Eq:SimpleGeomFunction}) (upper solid
line) and light collection function (\ref{Eq:SimpleGeomFuncWithAbs})
with absorption length $L_{A}=12$ m (lower solid line). One can see
that the equations (\ref{DeltaX}) and (\ref{DeltaY}) describe very
well the precision of reconstruction when using the light collection
function obtained from the experimental data. The results obtained
with the simplified geometrical function describe well only the resolution
at the detector's center. The influence of light refraction at the
scintillator/water interface significantly changes the light collection
function for events far away from the center, leading to the better
resolution at the detector's outer region.

\subsection{Spatial resolution of the CTF and Borexino detectors}

At the end let us give some examples of the spatial reconstruction
estimation. For the CTF detector $L(0)=L_{0}$ is 275 cm, the light
yield $A=360$ p.e./MeV, the averaged relative variance of the single
photoelectron response $v_{1}^{Det}=0.34$ (from the PMT test data).
The energy resolution is $R_{0}(250keV)=0.122$. The theoretical resolution
calculated for 100 PMTs using formula (\ref{formula:SigXq}) at $L_{A}=12$$\: $m
gives $\sigma _{x}^{(q)}\approx 37$ cm, which is almost two times
worse than the resolution that can be obtained using time signals
only. 

For the Borexino $L_{0}=657$ cm, the light yield $A=400$ p.e./MeV,
$v_{1}^{Det}=0.6$. The resolution calculated using (\ref{formula:SigXq})
at $L_{A}=12$$\: $m is $\sigma _{x}^{(q)}\approx 75$ cm, which
is much worse than the resolution available with the method based
on time signals.

\begin{figure*}

\caption{\label{Fig:SpatialBorexino}The precision of the spatial reconstruction
using charge signals for Borexino as a function of the distance from
the source to the detector's center. The simple light collection function
with absorption length of 12 meters has been used in estimations.}

\begin{center}\includegraphics[  width=1.0\paperwidth,
  height=0.50\paperwidth,
  keepaspectratio]{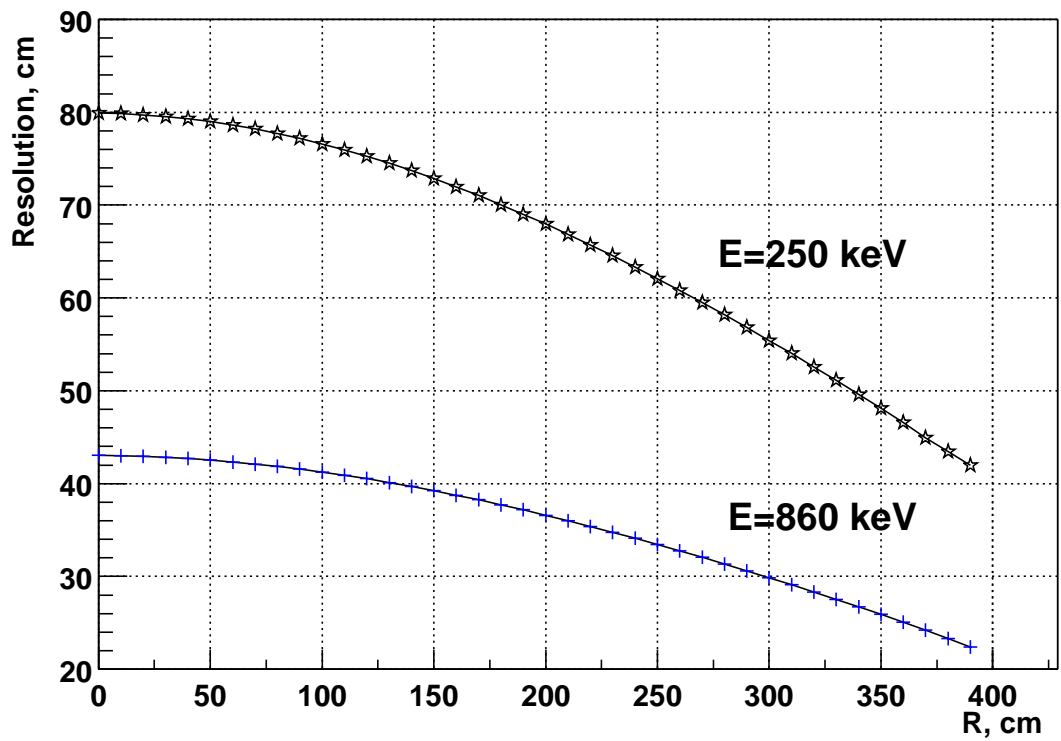}\end{center}
\end{figure*}

Calculations of the precision of coordinates reconstruction using
charge signals are presented in fig.\ref{Fig:SpatialBorexino}.$L_{A}=12$$\: $m
for two source energies (250 and 862 keV).

\subsection{Method of reconstruction using time signals}

If one of the PMTs registers an event at time $t_{0}$, and the $i$--th
PMT registers the same event at the time $t_{i}$, then:

\textbf{\[
t_{0}=T_{0}+tof_{0}+tt_{0}+\tau _{0}\, ,\]
}

\textbf{\[
t_{i}=T_{0}+tof_{i}+tt_{i}+\tau _{i}\, ,\]
}

where $T_{0}$ is the time at which the event occurred and $tof_{i}$
is the minimum time of flight for the photon from the event position
$\{x,y,z\}$ to the $i$--th PMT. The drift time of the electrons
inside the $i$--th PMT is $tt_{i}$, and $\tau _{i}$ is the moment
when the first photon, registered at the $i$--th PMT, has been emitted.
Index {}``0'' is used for the arbitrary PMT (the first one satisfying
$T_{min}<t_{i}<T_{max}$ condition). The time of registering of the
first photon by the $i$--th PMT can be calculated from the time of
arrival of the first photon at one of the PMTs:

\textbf{\begin{equation}
\tau _{i}=\tau _{0}+(tof_{0}+tt_{0}-t_{0})-(tof_{i}+tt_{i}-t_{i})\, .\label{tau1}\end{equation}
}

The reconstruction of an event position can be performed using the
maximum likelihood method. The 3 coordinates and one timing parameter
\textbf{$\tau _{0}$} are free parameters. The total charge $Q_{0}$,
corresponding to the event of the same energy at the detector's center,
is fixed. The likelihood function can be written as: \textbf{}

\textbf{\begin{equation}
L(x,y,z,\tau _{0})=\ln \left(\prod _{T_{min}<t_{i}<T_{max_{i}}}^{i=1..N_{PM}}p\, \left(\tau (\overrightarrow{r_{i}}(x,y,z),\tau _{0},t_{i}),\, \mu (Q_{0},\overrightarrow{r_{i}}(x,y,z)),\, p_{t}\right)\right)\, ,\label{MakLikeTime}\end{equation}
}

where 

\begin{lyxlist}{00.00.0000}
\item [$p\, \left(\tau (\overrightarrow{r_{i}}(x,y,z),\tau _{0},t_{i}),\, \mu (Q_{0},\overrightarrow{r_{i}}(x,y,z),\, p_{t})\right)$]is
the probability density function to observe the first pulse on the
$i$--th PMT at time $\tau $, for an event with coordinates $\{x,y,z\}$
in the detector's coordinate system, if the first photon at the $i$--th
PMT has been registered at the time $t_{i}$;
\item [$\tau \, (\overrightarrow{r_{i}}(x,y,z),\tau _{0},t_{i})$]is the
function that gives the time when the first photon, registered at
the $i$--th PMT, has been emitted. It takes into account the time
of flight of the photon from the position with coordinates $\overrightarrow{r_{i}}(x,y,z)$
in the $i$--th PMT coordinates system to the $i$--th PMT; 
\item [$\mu \left(Q_{0},\overrightarrow{r_{i}}(x,y,z)\right)$]is the mean
charge registered at the $i$--th PMT for an event at the position
with coordinates $\overrightarrow{r_{i}}(x,y,z)$ and the energy that
corresponds to the $Q_{0}$ total charge registered for an event of
the same energy at the detector's center; 
\item [$Q_{0}=\frac{Q}{f_{s}(x,y,z)}$]is the total charge registered for
an event of the same energy at the detector's center; 
\item [$\tau _{0}$]is a free timing parameter. The parameter $\tau _{0}$
coincides with $\tau \, \left(\overrightarrow{r_{i}}(x,y,z),\tau _{0}\right)$
for the first PMT which satisfies the relation $T_{min}<t_{i}<T_{max}$;
\item [$p_{t}$]is the part of the single electron response (SER) that
remains unregistered (discriminator threshold effect). It gives a
renormalization factor for the p.d.f.
\item [$T_{min}$]the lower limit on the time of PMT signal registering,
used for the rejection of random signals and false early triggering
of the electronics;
\item [$T_{max_{i}}$]is a hardware or software cut on the time registration
at the $i$--th PMT (whichever cut is smaller), applied in order to
improve the spatial reconstruction precision; 
\item [$\overrightarrow{r}_{i}(x,y,z)$]is the event coordinate in the
$i$--th PMT coordinate system;
\item [$(x,y,z)$]are the event coordinates in the detector coordinate
system. 
\end{lyxlist}
The algorithm of the likelihood function construction can be divided
into the following steps:

\begin{enumerate}
\item The initial total charge value $Q$ is fixed to the sum of the charge
registered at the individual PMTs. 
\item The initial coordinates $(x,y,z)$ of the event are guessed on the
basis of the signal distribution symmetries. As a first approximation
we assume the linear combination of the PMTs coordinates, with weights
corresponding to the time of the PMT signal arrival in respect to
the most probable time of arrival $t_{mp}$ :\[
x_{t}=-\frac{1}{2.24}\sum _{t<t_{mp}+2.5\sigma _{p}}^{i=1..N_{PM}}\frac{(t_{i}-t_{mp})x_{i}}{L_{0}}.\]
The most probable time of arrival $t_{mp}$ was defined in the following
way: first, the mean $<t>$ and variance $\sigma _{t}$ of the time
of arrival was defined over all the fired PMTs, then the position
of the peak $t_{p}$ is localized in the region $<t>\pm 3\sigma _{t}$
(as an approximation, the mean value in this region is used). The
width of the peak $\sigma _{p}$ is the variance of the time arrival
distribution in the same region. The coefficient is defined from the
measurements with the source at position \{0,0,94\}.
\item The charge $Q_{0}$ that corresponds to the event of the same energy
at the detector's center is calculated as $Q_{0}=\frac{Q}{f_{s}(x,y,z)};$
\item The condition $T_{min}<t_{i}<T_{max}$  is checked. If this condition
is false, the corresponding PMT is excluded from the maximum likelihood
calculation. 
\item The mean charge $\mu (\overrightarrow{r_{i}},Q_{0})$ expected at
the $i$--th PMT is defined using (\ref{Eq:mu(r,qu)});
\item For the first PMT that arrives at this point, the moment of time at
which the first photon has been emitted is assumed to be $\tau _{0}$,
and the parameter $T_{0}=tof_{0}+tt_{0}-t_{0}$ is calculated (see
formula$\: $(\ref{tau1})). For all other PMTs the parameter $\tau _{i}$
is calculated as \mbox{\textbf{$\tau _{i}=\tau _{0}+T_{0}-(tof_{i}+tt_{i}-t_{i})$}};
\item Now the cut time at the $i$--th PMT $T_{cut_{i}}$ is calculated
using formula (\ref{T_i}) of the next section;
\item Then the probability of a PMT hit at the moment $\tau _{i}$ is calculated:
$p_{i}=\rho (\tau _{i},\mu _{i},p_{t})$;
\item The value of the likelihood function is increased by $ln\, p_{i}$,
and the algorithm is repeated for the next PMT starting from point
4.
\end{enumerate}
The p.d.f. (probability density function) of the registration time
of the first photon $\rho _{1}(t)$ has been studied in laboratory
conditions \cite{ScintProp}. The conditions were set in such a way
that a PMT was registering practically a single p.e. (the mean number
of the registered p.e. were about 0.05) with p.d.f. $\rho (t)\simeq \rho _{1}(t)$.
The time of the scintillation occurrences were measured by another
PMT with a high precision. It is easy to show that the pdf of the
registration time $t$ for the light pulse with the mean p.e. number
$\mu $ is

\textbf{\begin{equation}
\rho (t)=\frac{\mu \cdot \rho _{1}(t)}{1-e^{-\mu }}e^{-\mu F(t)}.\label{RhoT}\end{equation}
}

where $F(t)=\int _{-\infty }^{t}\rho _{1}(t)dt$. 

In order to take into account the transit time of the PMT $\rho _{TT}(t)$
in (\ref{RhoT}), it is necessary to replace : $\rho _{1}(t)\rightarrow \rho _{1}(t)\otimes \rho _{TT}(t)$
(the sign $\otimes $ is used for the convolution of two functions).

In step 8 the properly normalized probability density function should
be used, i.e. $\rho (t)\rightarrow \frac{\rho (t)}{\int _{-\infty }^{T_{Cut_{i}}}\rho (t)dt}$.
But for a proper treatment of the likelihood function, the conditional
probability for the first hit to occur within the $[-\infty ;T_{Cut_{i}}]$
interval should be multiplied later by the probability of the channel
hit within $[-\infty ;T_{Cut_{i}}]$. The last observation permits
skipping the unnecessary calculation of the $\int _{-\infty }^{t}\rho _{1}(t)dt$
integral. 

It is important to notice that it was assumed that the p.d.f. of the
time of the registering of the first photon is independent of the
source position.

\subsection{Analysis of the precision of spatial reconstruction using time data}

When reconstructing an event position using (\ref{MakLikeTime}) one
should note that late registered photons do not provide information
about the event coordinates, so such signals should be excluded from
the analysis. The influence of the different {}``time cuts'' on
the reconstruction precision is investigated below. The $T_{cut}$
\textbf{}is counted from the moment \textbf{$(T_{0}+tof_{min})$}
where \textbf{$T_{0}$} is an event occurrence time, and \textbf{$tof_{min}$}
is time of flight to the closest PMT in the detector.

The uncertainty of the time of arrival of the photon to a single PMT
is:

\textbf{\begin{equation}
\sigma _{t}(T)=\frac{\sqrt{\sigma ^{2}(T)+\sigma _{T_{0}}^{2}(T)}}{1-e^{-\mu F(T)}},\label{Eq:unc1}\end{equation}
} where \textbf{\begin{equation}
\sigma ^{2}(T)=\frac{\int _{-\infty }^{T}\left(t-\overline{t(T)}\right)^{2}\rho (T)dt}{F(T)},\label{Eq:SigmaSqT}\end{equation}
} and \textbf{\begin{equation}
\sigma _{T_{o}}^{2}=\left(\sum _{i}^{N_{PM}}\frac{1-e^{-\mu F(T_{i})}}{\sigma _{i}^{2}(T)}\right)^{-1}\: \label{Eq:SigmaSqT_0}\end{equation}
} is the uncertainty of the reconstruction of $T_{0}$ for an event
with known coordinates. The denominator $1-e^{-\mu F(t)}$ reflects
the fact that the photon is registered in the time interval $[-\infty ,T]$.
The {}``cut time'' for the $i$--th PMT is:

\textbf{\begin{equation}
T_{cut_{i}}=T_{cut}+T_{0}+tof_{min}-tof_{i}\: .\label{T_i}\end{equation}
}

So, the closer to a PMT the event occurs, the bigger is the {}``cut
time''.

Using simple geometrical relations (see fig.\ref{fGeom}) one can
write:

\textbf{\begin{equation}
L(r,\Theta )=\sqrt{L_{0}^{2}+r^{2}-2\cdot r\cdot L_{0}\cdot \cos (\Theta )}\: ,\label{Formula:8}\end{equation}
}

so \textbf{\begin{equation}
\frac{dL}{dr}=\frac{r-L_{0}\cdot \cos (\Theta )}{L(r,\Theta )}\label{Formula:9}\end{equation}
} and \textbf{\begin{equation}
\frac{dL}{d\cos \Theta }=-\frac{r\cdot L_{0}\cdot \cos (\Theta )}{L(r,\Theta )}.\label{Formula:10}\end{equation}
}

A small change of the source position by $\Delta r$ along the radius
can be registered by a PMT if \textbf{\mbox{$\frac{dL(r,\cos (\Theta ))}{dr}\Delta r\approx \sigma _{t}\cdot \frac{c}{n}$
}} from where: 

\textbf{\begin{equation}
\sigma _{r}=\frac{c}{n}\frac{\sigma _{t}(T)}{\frac{dL(r,y)}{dr}}.\label{Formula:11}\end{equation}
}

Summing over all PMTs and substituting the summing with an integration
over \mbox{\textbf{$y\equiv \cos (\Theta )$}}\textbf{,} we will
obtain

\textbf{\begin{equation}
\sigma _{T_{0}}^{2}=\frac{1}{N_{PM}}\left(\frac{1}{2}\int _{-1}^{+1}\frac{1-e^{-\mu F(T(r,y))}}{\sigma ^{2}\left(T(r,y)\right)}dy\right)^{-1},\label{SigmaT0}\end{equation}
}

and

\begin{equation}
(\frac{1}{\sigma _{r}})^{2}=\sum \frac{1}{\sigma _{r_{i}^{2}}}\approx N_{PM}\cdot \frac{1}{2}\int _{-1}^{+1}\frac{\left(\frac{dL(r,y)}{dr}\right)^{2}}{(\frac{c}{n})^{2}\sigma _{t}^{2}\left(T(r,y)\right)}dy\: .\label{Formula:12}\end{equation}

Taking into account the relations for \textbf{$\frac{dL}{dr}$} and
\textbf{$\sigma _{t}(T)$} we can finally write:

\begin{equation}
\sigma _{r}(r)=\frac{1}{\sqrt{N_{PM}}}\frac{c}{n}\left[\frac{1}{2}\int _{-1}^{+1}\left(\frac{r-L_{0}\cdot y}{L(r,y)\cdot \sigma _{t}\left(T(r,y)\right)}\right)^{2}\left(1-e^{-\mu (r,y)F\left(T(r,y)\right)}\right)dy\right]^{-1}.\label{SigmaR}\end{equation}

The same relation can be obtained for the tangential resolution:

\begin{equation}
\left(\frac{1}{\sigma _{\Theta }}\right)^{2}=\sum \frac{1}{\sigma _{\Theta _{i}^{2}}}\approx N_{PM}\cdot \frac{1}{2}\int _{-1}^{+1}\frac{(\frac{1}{r}\frac{dL(r,y)}{dy})^{2}}{(\frac{c}{n})^{2}\sigma _{t}^{2}\left(T(r,y)\right)}dy\: ,\label{Formula:13}\end{equation}

and

\begin{equation}
\sigma _{\Theta }(r)=\frac{1}{\sqrt{N_{PM}}}\frac{c}{n}\left[\frac{1}{2}\int _{-1}^{+1}\left(\frac{L_{0}\cdot y}{L(r,y)\cdot \sigma (T)}\right)^{2}\left(1-e^{-\mu (r,y)F\left(T(r,y)\right)}\right)dy\right]^{-1}.\label{SigmaTheta}\end{equation}

It should be noted that the estimate for the mean time of flight is
only approximate due to the simplified geometrical relations used.
In the CTF detector for events close to the inner vessel, the refraction
effects at the scintillator/water interface will complicate the precise
time of flight estimation. Nevertheless, comparison with a precise
calculation shows that these effects can be neglected.

\subsection{Precision of spatial reconstruction using time data for the case
of an event at the detector's center.}

The precision of spatial reconstruction using time data for the events
at the detector's center follows from formula (\ref{SigmaR}) for
$r=0$: \textbf{}

\begin{equation}
\sigma _{x}^{(t)}=\sqrt{3}\frac{\frac{c}{n}\sqrt{\sigma ^{2}(T)+\sigma _{T_{0}}^{2}(T)}}{\sqrt{N_{PM}\left(1-e^{-\mu _{0}F(T)}\right)}},\label{SigmaXteq}\end{equation}

where 

\begin{eqnarray*}
\sigma (T) & = & \frac{\int _{T_{min}}^{T}\rho (t)\left(t-<t(T)>\right)^{2}dt}{F(T)}\\
<t(T)> & = & \frac{\int _{T_{min}}^{T}\rho (t)tdt}{F(T)},\\
F(T) & = & \int _{T_{min}}^{T}\rho (t)dt\, ,
\end{eqnarray*}

and:

\begin{lyxlist}{00.00.0000}
\item [\textbf{n}]is the scintillator refraction index;
\item [\textbf{$N_{PM}\left(1-e^{-\mu _{0}F(T)}\right)$}]is the mean number
of PMTs triggered in the time interval \textbf{{[}$T_{min};T${]}}
for an event with the mean number of photoelectrons registered per
PMT $\mu _{0}$;
\item [\textbf{$T_{0}$}]is the moment of scintillation;
\end{lyxlist}
The $T_{min}$ parameter here (\textbf{$T_{min}<T_{0}$}) is chosen
to satisfy the relation \textbf{\mbox{$\int _{-\infty }^{T_{min}}\rho (t)dt\simeq 0$}.}
The factor $\sqrt{3}$ - appears from the averaging over all PMTs
(volume factor).

\subsection{Precision of the spatial reconstruction using time data calculated
for CTF and Borexino.}

\begin{figure*}

\caption{\label{Fig:CTF_tcut}The precision of the spatial reconstruction
for CTF as a function of \textbf{$T_{cut}$} for the source energy
250 keV. The time data only are used in estimations. Two \textbf{}sources
with different energies have been considered at different source distances
from the detector's center.}

\begin{center}\includegraphics[  width=0.90\paperwidth,
  height=0.40\paperwidth,
  keepaspectratio]{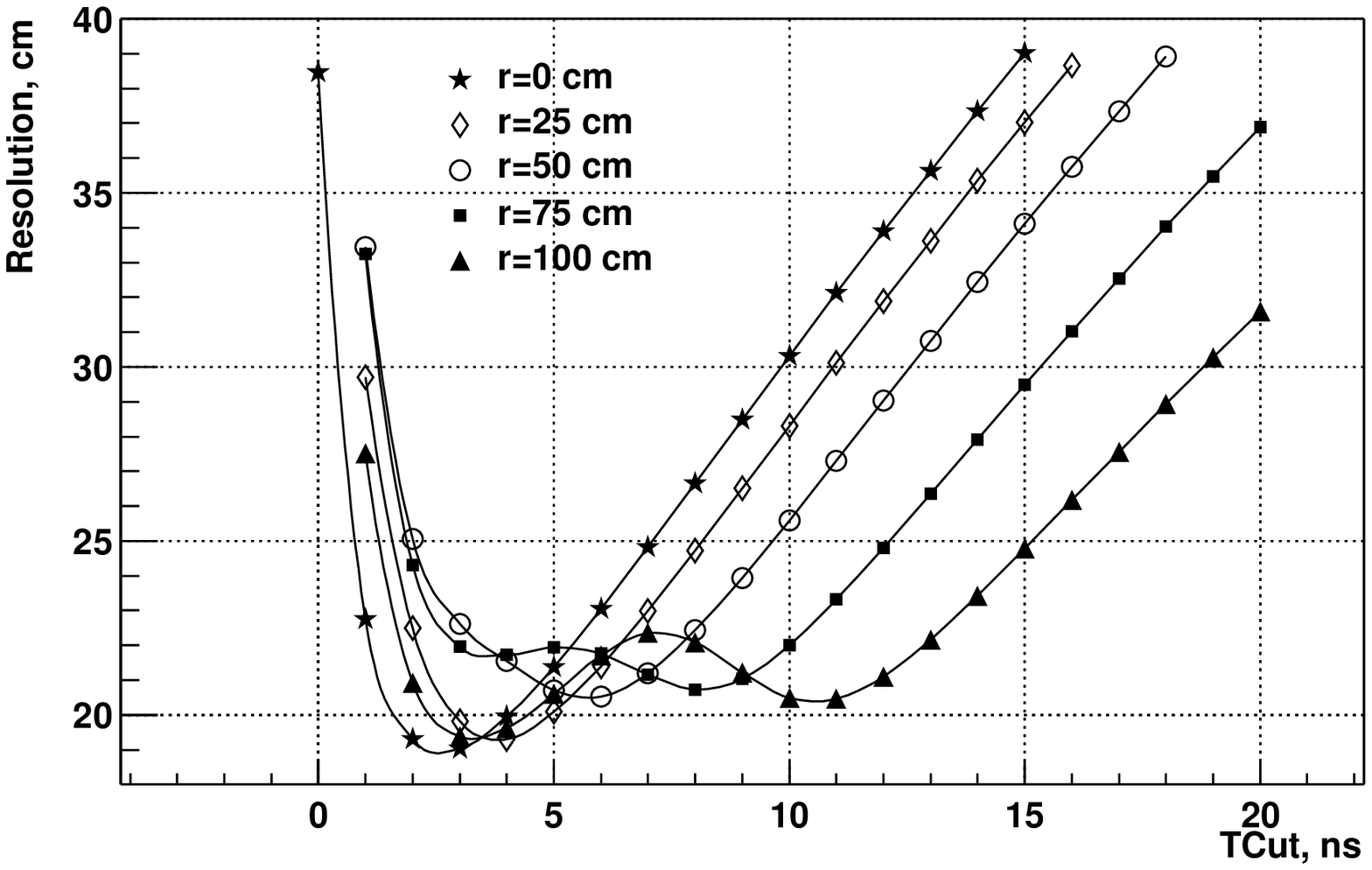}\end{center}\vspace{0.375cm}

\end{figure*}

The estimate of the precision of the spatial reconstruction for an
event of 250 keV energy in CTF \textbf{}is presented in fig.\ref{Fig:CTF_tcut}
as a function of \textbf{$T_{cut}$}. The light yield is $A=360$
p.e./MeV (for 100 PMTs). 

\begin{figure*}

\caption{\label{Fig:Borexino_tcut}The precision of the spatial reconstruction
for Borexino as a function of \textbf{$T_{cut}$} for the source energy
250 keV. The time data only are used in estimations. Two \textbf{}sources
with different energies have been considered at different source distances
from the detector's center.}

\begin{center}\includegraphics[  width=0.90\paperwidth,
  height=0.40\paperwidth,
  keepaspectratio]{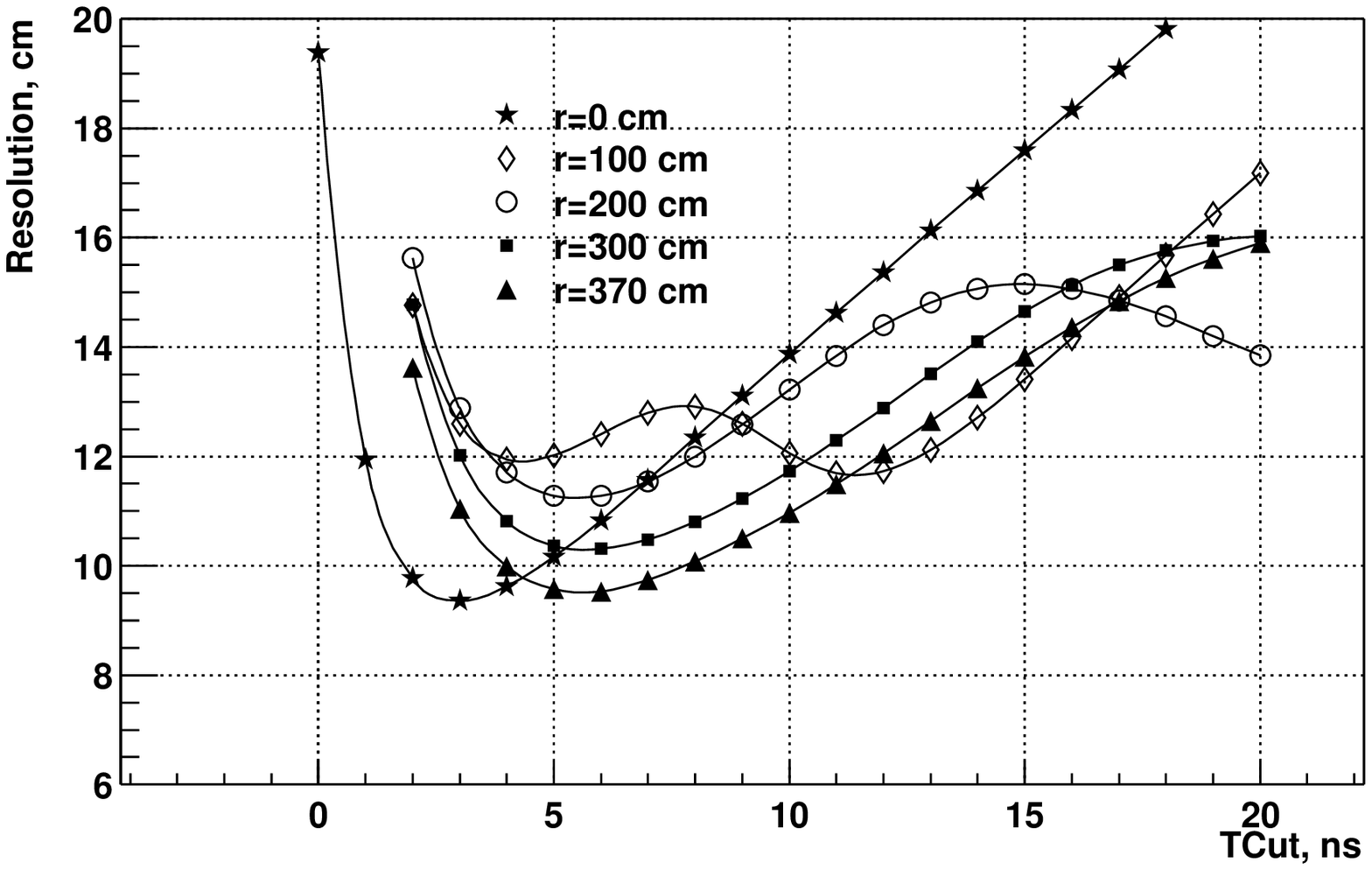}\end{center}
\end{figure*}

The precision of the spatial reconstruction for an event of 250 keV
energy in Borexino calculated using (\ref{SigmaR}) \textbf{}and (\ref{SigmaTheta})
is presented in fig.\ref{Fig:Borexino_tcut} as a function of \textbf{$T_{cut}$}.
Here, the light yield is $A=400$ p.e./MeV (for 2200 PMTs). The zero
of the time scale corresponds to the scintillation event at the $T_{0}$.
The best resolution is achieved for a $T_{cut}$ of about 3-5 ns,
then the resolution degrades, inspite of the fact that the total amount
of signals increases. It should be pointed out that the resolution
is overestimated, since the achievable resolution will be worse because
of broadening of the real p.d.f. of the time of arrival of the first
p.e., in comparison to the p.d.f obtained in the laboratory measurements
on a small volumes of scintillator.

\subsection{\label{Sec:Q+T_reconstruction}Method of reconstruction using time
and charge signals simultaneously}

The reconstruction of an event position is performed using the maximum
likelihood method with 5 free parameters: 3 coordinates, one timing
parameter \textbf{$\tau _{0}$,} and the total charge \textbf{$Q_{0}$},
corresponding to an event of the same energy at the detector's center.
The likelihood function is in this case the sum of the likelihood
functions (\ref{Eq:MaxLikeCharge}) and (\ref{MakLikeTime}):

$L(x,y,z,\tau _{0},Q_{0})=$

$=\ln \left(\prod _{i=1,t_{i}<T_{max_{i}}}^{N_{PM}}p\, \left(\tau \left(\overrightarrow{r_{i}}(x,y,z),\tau _{0},t_{i})\right),\, \mu \left(Q_{0},\overrightarrow{r_{i}}(x,y,z)),\, p_{t}\right)\right)\right)+$

\textbf{\begin{equation}
+\ln \left(\prod _{i=1}^{N_{PM}}p\; \left(\mu (\overrightarrow{r_{i}}(x,y,z),Q_{0}),q_{i}\right)\right)\, .\label{Eq:MaxLikelyhoodQT}\end{equation}
}

The notations used are the same as in (\ref{Eq:MaxLikeCharge}) and
(\ref{MakLikeTime}). 

One can expect that the precision of reconstruction using (\ref{Eq:MaxLikelyhoodQT})
will be better than the separate resolutions for the reconstruction
using the time signals and the one using charge signals.

\begin{figure*}

\caption{\label{Fig:CTF_qt_tcut}The precision of the spatial reconstruction
precision for CTF as a function of $T_{cut}\, $for the source energy
250 keV.The time and charge data are used in estimations. Two \textbf{}sources
with different energies have been considered at different source distances
from the detector's center.}

\begin{center}\includegraphics[  width=0.90\paperwidth,
  height=0.40\paperwidth,
  keepaspectratio]{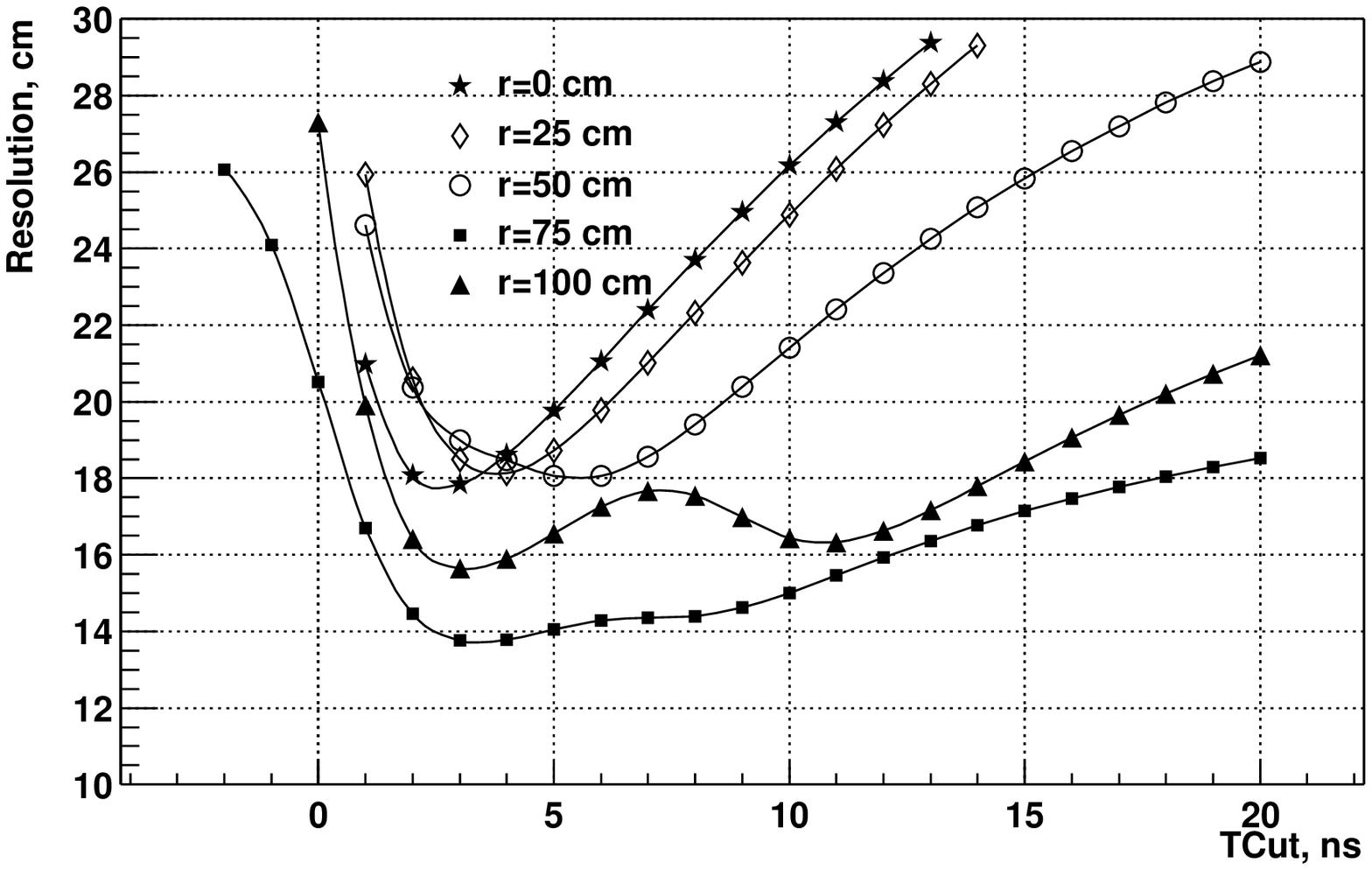}\end{center}
\end{figure*}

\begin{figure*}

\caption{\label{Fig:Borexino_qt_tcut}The precision of the spatial reconstruction
for Borexino using charge and time data as a function of $T_{cut}\, $for
the source energy 250 keV. The time and charge data are used in estimations.
Two \textbf{}sources with different energies has been considered at
different source distances from the detector's center.}

\begin{center}\includegraphics[  width=0.90\paperwidth,
  height=0.40\paperwidth,
  keepaspectratio]{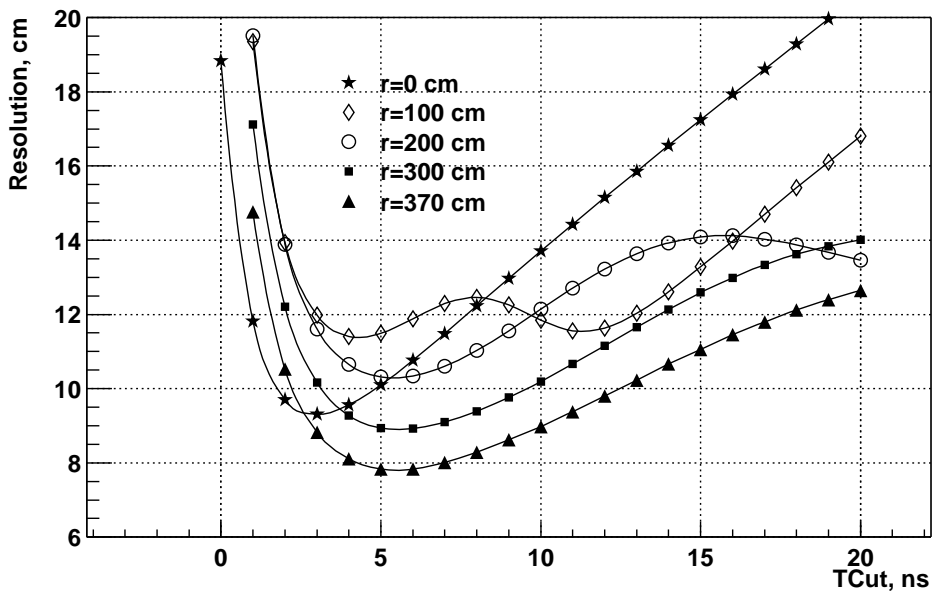}\end{center}
\end{figure*}

\begin{figure*}

\caption{\label{Fig:CTF_improvement}The result of reconstruction of the source
position in CTF relative to the nominal source position ($dR$ is
the distance between the reconstructed source position and the nominal
one) using the time and charge data (dotted line). The artificial
radon source is at the position r(4,-63,76)=98 cm from the detector's
center. The distance dR from the reconstructed source position to
the nominal one is presented. The plot shows also the reconstruction
using the time data only (solid line).}

\begin{center}\includegraphics[  width=1.0\paperwidth,
  height=0.40\paperwidth,
  keepaspectratio]{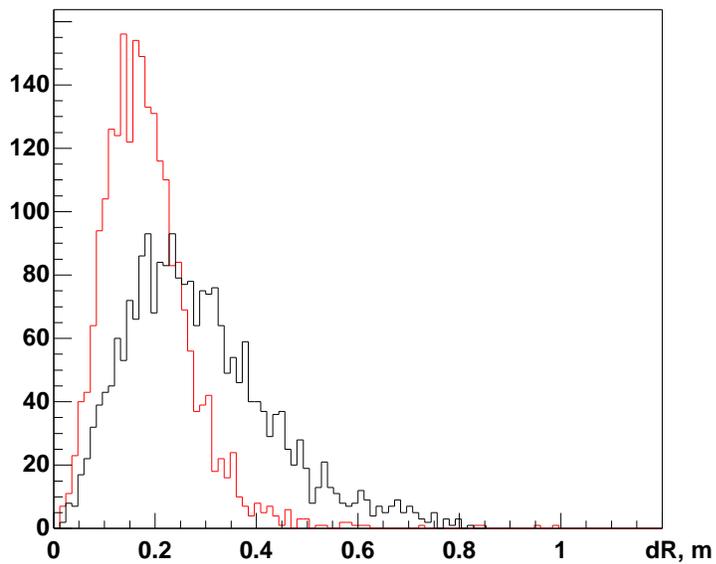}\end{center}
\end{figure*}

The precision of the spatial reconstruction for CTF \textbf{}and Borexino
are presented in fig.\ref{Fig:CTF_qt_tcut} and \ref{Fig:Borexino_qt_tcut}
as a function of \textbf{$T_{cut}$}. These plots should be compared
to those in fig.\ref{Fig:CTF_tcut} and fig.\ref{Fig:Borexino_tcut}.
One can see that significant improvement of the resolution can be
obtained, especially in the region near the inner vessel boundary.
In fig.\ref{Fig:CTF_improvement} the results of the reconstruction
using the likelihood function (\ref{Eq:MaxLikelyhoodQT}) for a source
near the inner vessel boundary are presented; one can see significant
improvement of the spatial resolution.

\begin{figure*}

\caption{\label{Fig:Energy} The result of the combined reconstruction (using
time and charge data) are shown with circles. The total charge registered
for the different source positions, defined as a sum over all PMTs,
is shown for comparison (marked by crosses). The reconstruction results
are renormalized in order to make the comparison more evident (the
point at $r\sim 0$ cm has been used for renormalization). One can
see that reconstruction program provides a better base value for the
energy evaluation.}

\begin{center}\includegraphics[  width=1.0\paperwidth,
  height=0.50\paperwidth,
  keepaspectratio]{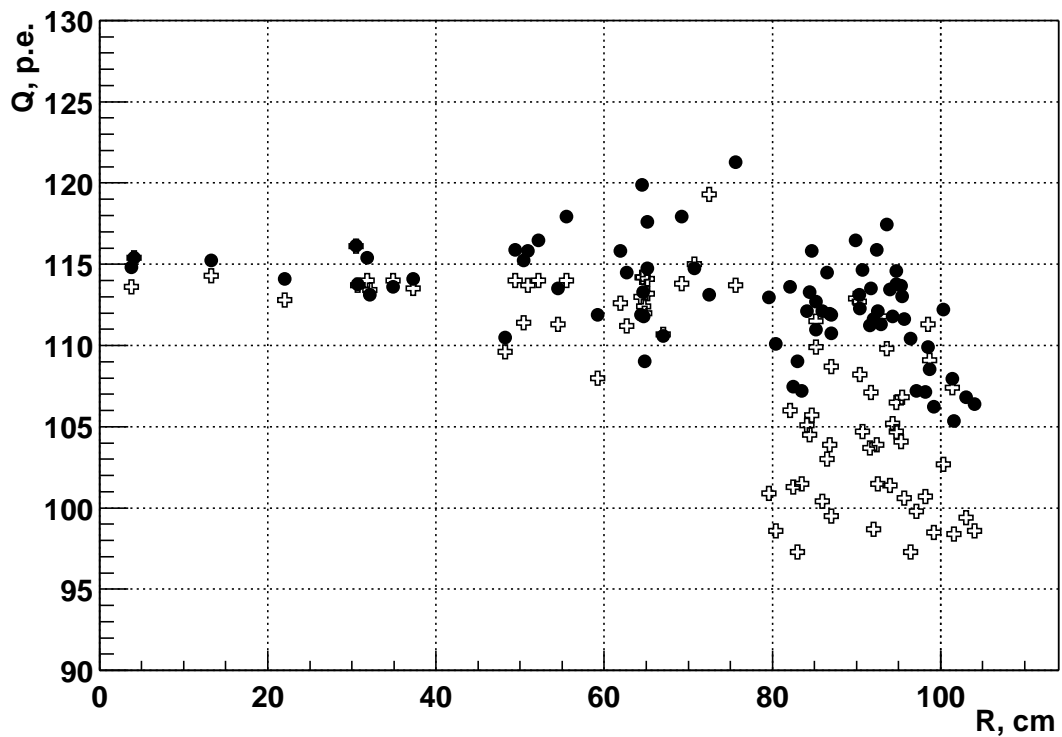}\end{center}
\end{figure*}

The results of the source energy reconstruction for CTF, maximizing
the likelihood function (\ref{Eq:MaxLikelyhoodQT}) are presented
in fig.\ref{Fig:Energy}. The mean value and its variance for the
reconstructed energy over different source positions is \textbf{$137.9\pm 3.9\, (2.8\%)\: p.e.$}
One can notice that the reconstruction with (\ref{Eq:MaxLikelyhoodQT})
provides a better base value for the reconstruction of energy (see
section \ref{SubSection:CTFlightCollection}). The larger absolute
value of the mean charge in comparison to the values from section
\ref{SubSection:CTFlightCollection} is the result of a different
PMT calibration procedure used in the reconstruction program and that
used in the standard CTF calibration procedure. In the standard CTF
reconstruction program the position of the SER is taken as the calibration
value (i.e. the most probable value). The reconstruction with the
PMT charge signals needs a more precise calibration of the PMTs using
the mean value of the SER (see \cite{ScintProp2}). For a typical
PMT the position of the mean is $15-20\%$ lower that the position
of the peak (or the most probable value). This difference has no influence
on the detector energy resolution when calculating the energy from
the registered charge using (3) (see section \ref{Sec:CalibrationInfluence}
for the explanation). The only noticeable change is the energy scale
factor, used to transform p.e. to energy.

\section{Conclusions}

The energy and spatial resolutions of a large volume liquid scintillator
detector with a spherical symmetry are discussed in detail. An event
reconstruction technique using charge and time data from the PMTs
is analyzed in order to obtain optimal detector resolutions. The relations
for the numerical estimations of the energy and spatial resolutions
are obtained and verified with the CTF detector data.

\section*{Acknowledgements}

This job was performed with a support of the INFN Milano section in
accordance to the scientific agreement on Borexino between INFN and
JINR (Dubna). I would like especially to thank Prof. G. Bellini and
Dr. G. Ranucci who organized my stay at the LNGS laboratory, and to
E. Meroni for the continuous interest in the reconstruction program.

I would also like to thank the following people: Dr. G. Ranucci who
was the first to point out to me the possibility to use charge signals
from the PMTs for the spatial reconstruction. Prof. O. Zaimidoroga
for useful discussions and support. Special thanks to all my colleagues
from the Borexino collaboration, especially to A. Ianni, G. Korga,
L. Papp. I am very grateful to R. Ford and V. Kobychev for the careful
reading of the manuscript and really useful discussions.

\newpage
\listoftables

\newpage
\listoffigures

\newpage

\begin{table*}

\caption{\label{Table:Radon}Energy resolution for the radon dissolved in
the scintillator volume.}

$ $

$ $

$ $

\begin{tabular}{|c|c|c|c|c|c|}
\hline 
\selectlanguage{english}
r, cm\selectlanguage{american}
&
\multicolumn{2}{c|}{\selectlanguage{english}
 Data \selectlanguage{american}
}&
\multicolumn{2}{c|}{\selectlanguage{english}
Gauss fit \selectlanguage{american}
}&
\selectlanguage{english}
$N_{event}$\selectlanguage{american}
\\
\cline{2-3} \cline{4-5} 
&
\selectlanguage{english}
<Q>\selectlanguage{american}
&
\selectlanguage{english}
$\sigma _{Q}$\selectlanguage{american}
&
\selectlanguage{english}
<Q>\selectlanguage{american}
&
\selectlanguage{english}
$\sigma _{Q}$\selectlanguage{american}
&
\\
\hline 
\selectlanguage{english}
$0<r<10$\selectlanguage{american}
&
\selectlanguage{english}
238.5\selectlanguage{american}
&
\selectlanguage{english}
12.8\selectlanguage{american}
&
\selectlanguage{english}
---\selectlanguage{american}
&
\selectlanguage{english}
---\selectlanguage{american}
&
\selectlanguage{english}
9\selectlanguage{american}
\\
\hline 
\selectlanguage{english}
$10<r<20$\selectlanguage{american}
&
\selectlanguage{english}
246.5\selectlanguage{american}
&
\selectlanguage{english}
17.2\selectlanguage{american}
&
\selectlanguage{english}
---\selectlanguage{american}
&
\selectlanguage{english}
---\selectlanguage{american}
&
\selectlanguage{english}
58\selectlanguage{american}
\\
\hline 
\selectlanguage{english}
$20<r<30$\selectlanguage{american}
&
\selectlanguage{english}
243.6\selectlanguage{american}
&
\selectlanguage{english}
19.6\selectlanguage{american}
&
\selectlanguage{english}
---\selectlanguage{american}
&
\selectlanguage{english}
---\selectlanguage{american}
&
\selectlanguage{english}
157\selectlanguage{american}
\\
\hline 
\selectlanguage{english}
$30<r<40$\selectlanguage{american}
&
\selectlanguage{english}
243.7\selectlanguage{american}
&
\selectlanguage{english}
17.5\selectlanguage{american}
&
\selectlanguage{english}
244.4\selectlanguage{american}
&
\selectlanguage{english}
18.0\selectlanguage{american}
&
\selectlanguage{english}
302\selectlanguage{american}
\\
\hline 
\selectlanguage{english}
$40<r<50$\selectlanguage{american}
&
\selectlanguage{english}
245.4\selectlanguage{american}
&
\selectlanguage{english}
19.2\selectlanguage{american}
&
\selectlanguage{english}
246.0\selectlanguage{american}
&
\selectlanguage{english}
20.4\selectlanguage{american}
&
\selectlanguage{english}
505\selectlanguage{american}
\\
\hline 
\selectlanguage{english}
$50<r<60$\selectlanguage{american}
&
\selectlanguage{english}
244.6\selectlanguage{american}
&
\selectlanguage{english}
19.1\selectlanguage{american}
&
\selectlanguage{english}
244.7\selectlanguage{american}
&
\selectlanguage{english}
18.0\selectlanguage{american}
&
\selectlanguage{english}
657\selectlanguage{american}
\\
\hline 
\selectlanguage{english}
$60<r<70$\selectlanguage{american}
&
\selectlanguage{english}
244.2\selectlanguage{american}
&
\selectlanguage{english}
18.9\selectlanguage{american}
&
\selectlanguage{english}
244.9\selectlanguage{american}
&
\selectlanguage{english}
18.4\selectlanguage{american}
&
\selectlanguage{english}
972\selectlanguage{american}
\\
\hline 
\selectlanguage{english}
$70<r<80$\selectlanguage{american}
&
\selectlanguage{english}
241.0\selectlanguage{american}
&
\selectlanguage{english}
20.6\selectlanguage{american}
&
\selectlanguage{english}
240.6\selectlanguage{american}
&
\selectlanguage{english}
20.7\selectlanguage{american}
&
\selectlanguage{english}
1058\selectlanguage{american}
\\
\hline 
\selectlanguage{english}
$80<r<90$\selectlanguage{american}
&
\selectlanguage{english}
237.0\selectlanguage{american}
&
\selectlanguage{english}
21.5\selectlanguage{american}
&
\selectlanguage{english}
237.5\selectlanguage{american}
&
\selectlanguage{english}
21.5\selectlanguage{american}
&
\selectlanguage{english}
1168\selectlanguage{american}
\\
\hline 
\selectlanguage{english}
$90<r<100$\selectlanguage{american}
&
\selectlanguage{english}
233.3\selectlanguage{american}
&
\selectlanguage{english}
21.0\selectlanguage{american}
&
\selectlanguage{english}
233.4\selectlanguage{american}
&
\selectlanguage{english}
21.6\selectlanguage{american}
&
\selectlanguage{english}
995\selectlanguage{american}
\\
\hline 
\selectlanguage{english}
$100<r<110$\selectlanguage{american}
&
\selectlanguage{english}
229.9\selectlanguage{american}
&
\selectlanguage{english}
21.3\selectlanguage{american}
&
\selectlanguage{english}
229.7\selectlanguage{american}
&
\selectlanguage{english}
22.2\selectlanguage{american}
&
\selectlanguage{english}
508\selectlanguage{american}
\\
\hline 
\selectlanguage{english}
$110<r<\infty $\selectlanguage{american}
&
\selectlanguage{english}
231.5\selectlanguage{american}
&
\selectlanguage{english}
22.4\selectlanguage{american}
&
\selectlanguage{english}
230.6\selectlanguage{american}
&
\selectlanguage{english}
23.5\selectlanguage{american}
&
\selectlanguage{english}
285\selectlanguage{american}
\\
\hline 
\selectlanguage{english}
$0<r<30$\selectlanguage{american}
&
\selectlanguage{english}
244.2\selectlanguage{american}
&
\selectlanguage{english}
18.9\selectlanguage{american}
&
\selectlanguage{english}
---\selectlanguage{american}
&
\selectlanguage{english}
---\selectlanguage{american}
&
\selectlanguage{english}
224\selectlanguage{american}
\\
\hline 
\selectlanguage{english}
$0<r<40$\selectlanguage{american}
&
\selectlanguage{english}
243.9\selectlanguage{american}
&
\selectlanguage{english}
18.1\selectlanguage{american}
&
\selectlanguage{english}
244.4\selectlanguage{american}
&
\selectlanguage{english}
17.7\selectlanguage{american}
&
\selectlanguage{english}
526\selectlanguage{american}
\\
\hline 
\selectlanguage{english}
$0<r<50$\selectlanguage{american}
&
\selectlanguage{english}
244.6\selectlanguage{american}
&
\selectlanguage{english}
18.6\selectlanguage{american}
&
\selectlanguage{english}
245.2\selectlanguage{american}
&
\selectlanguage{english}
18.5\selectlanguage{american}
&
\selectlanguage{english}
1831\selectlanguage{american}
\\
\hline 
\selectlanguage{english}
$0<r<\infty $\selectlanguage{american}
&
\selectlanguage{english}
239.3\selectlanguage{american}
&
\selectlanguage{english}
20.9\selectlanguage{american}
&
\selectlanguage{english}
239.2\selectlanguage{american}
&
\selectlanguage{english}
21.1\selectlanguage{american}
&
\selectlanguage{english}
6674\selectlanguage{american}
\\
\hline
\end{tabular}
\end{table*}

\newpage



\end{document}